\documentclass{aa}
\usepackage{graphicx}
\usepackage{natbib}
\usepackage{txfonts}
\bibpunct{(}{)}{;}{a}{}{,}
\newcommand{\kpc}{\,{\rm kpc}}
\newcommand{\ha}{H$\alpha$}

\newcommand{\kms}{\,km\,s$^{-1}$}  
\newcommand{\myr}{\,$M_{\sun}\,{\rm yr}^{-1}$}
\newcommand{\ro}{\,$R_{\sun}$}
\newcommand{\mo}{\,$M_{\sun}$}
\newcommand{\lo}{\,$L_{\sun}$}
\newcommand{\cmt}{\,cm$^{-3}$}

\newcommand{\cmd}{\,cm$^{-2}$}
\newcommand{\cmdd}{\,cm$^{2}$}
\newcommand{\es}{$\rm\,erg\,s^{-1}$}
\newcommand{\ecs}{$\rm\,erg\,cm^{-2}\,s^{-1}$}
\newcommand{\ecsa}{$\rm\,erg\,cm^{-2}\,s^{-1}\,\AA^{-1}$}
\begin{document}

\title{Early evolution of the extraordinary Nova Del 2013 (V339~Del)
\thanks{Based on data collected by amateur astronomers}}

\author{A.~Skopal\inst{1}\thanks{Visiting Astronomer: 
                                 Astronomical Institute, Bamberg}
     \and H.~Drechsel  \inst{2}
     \and T.~Tarasova  \inst{3}
     \and T.~Kato      \inst{4}
     \and M.~Fujii     \inst{5}
     \and F.~Teyssier  \inst{6}
     \and O.~Garde     \inst{7}
     \and J.~Guarro    \inst{8}
     \and J.~Edlin     \inst{9}
     \and C.~Buil      \inst{10}
     \and D.~Antao     \inst{11}
     \and J.-N.~Terry  \inst{12}
     \and T.~Lemoult   \inst{13}
     \and S.~Charbonnel\inst{14}
     \and T.~Bohlsen   \inst{15}
     \and A.~Favaro    \inst{16}
     \and K.~Graham    \inst{17}
}
\institute{Astronomical Institute, Slovak Academy of Sciences, 
           059\,60 Tatransk\'{a} Lomnica, Slovakia 
\and
Dr. Karl Remeis-Observatory \& ECAP, Astronomical Institute,
Friedrich-Alexander-Universit\"at Erlangen-N\"urnberg,\\
Sternwartstra\ss e 7, 96049 Bamberg, Germany
\and
Crimean Astrophysical Observatory, 298409 Nauchny, Crimea, Russia
\and
Department of Astronomy, Kyoto University, Kitashirakawa-Oiwake-cho,
Sakyo-ku, Kyoto 606-8502, Japan
\and 
Fujii Kurosaki Observatory, 4500 Kurosaki, Tamashima, Kurashiki, 
Okayama 713-8126, Japan 
\and 
67 Rue Jacques Daviel, Rouen 76100, France
\and
Observatoire de la Tourbi\`ere, 38690 Chabons, France
\and
Balmes 2, 08784 PIERA, Barcelona, Spain
\and
1833 Bobwhite Dr. Ammon, Idaho,  USA 83401
\and
Castanet Tolosan Observatory, 6 Place Clemence Isaure, 
31320 Castanet Tolosan, France
\and
Lieu-dit Durfort, 81150 Fayssac, France
\and
6 rue Virgile, 42100 Saint-Etienne, France
\and
Chelles Observatory, 23 avenue h\'enin, 77500 Chelles, France
\and
Durtal Observatory, 6 rue des Glycines, 49430 Durtal, France
\and 
Mirranook  Observatory, Boorolong Rd Armidale, NSW, Australia 2350
\and 
Societe Astronomique de Bourgogne-Dijon, France 
\and 
23746 Schoolhouse Road, Manhattan, Illinois, USA 60442
}
\date{Received / Accepted}

\abstract
 {}
 {
We determine the temporal evolution of the luminosity 
($L_{\rm WD}$), radius ($R_{\rm WD}$) and effective temperature 
($T_{\rm eff}$) of the white dwarf (WD) pseudophotosphere of 
V339~Del from its discovery to around day 40. Another main 
objective was studying the ionization structure of the ejecta. 
}
 {
These aims were achieved by modelling the optical/near-IR spectral 
energy distribution (SED) using low-resolution spectroscopy 
(3500--9200\,\AA), $UBVR_{\rm C}I_{\rm C}$ and $JHKLM$ photometry. 
Important insights in the physical conditions of the ejecta were 
gained from an analysis of the evolution of the \ha\ and 
Raman-scattered 6825\,\AA\ \ion{O}{vi} line using medium-resolution 
spectroscopy ($R\sim 10000$). 
 }
 {During the fireball stage (Aug. 14.8--19.9, 2013), 
$T_{\rm eff}$ was in the range of 6000--12000\,K, $R_{\rm WD}$ 
was expanding non-uniformly in time from $\sim$\,66 to 
$\sim$\,300\,$(d/3\,{\rm kpc})$\ro, and $L_{\rm WD}$ was 
super-Eddington, but not constant. 
Its maximum of $\sim$\,9$\times 10^{38}\,(d/3\,{\rm kpc})^2$\es\ 
occurred around Aug. 16.0, at the maximum of $T_{\rm eff}$, 
half a day before the visual maximum. 
After the fireball stage, a large emission measure of 
$1.0-2.0\times 10^{62}\,(d/3\,{\rm kpc})^2$\cmt\ constrained 
the lower limit of $L_{\rm WD}$ to be well above the 
super-Eddington value. The mass of the ionized region was 
a few $\times 10^{-4}$\mo, and the mass-loss rate was decreasing 
from $\sim$\,5.7 (Aug.~22) to $\sim$\,0.71$\times 10^{-4}$\myr\ 
(Sept.~20). 
The evolution of the \ha\ line and mainly the transient 
emergence of the Raman-scattered \ion{O}{vi} 1032\,\AA\ line 
suggested a biconical ionization 
structure of the ejecta with a disk-like \ion{H}{i} region 
persisting around the WD until its total ionization, around 
day 40. On Sept.~20 (day 35), the model SED indicated a dust 
emission component in the spectrum. 
The dust was located beyond the \ion{H}{i} zone, where it was 
shielded from the hard, $\ga 10^5$\,K, radiation of the burning 
WD at that time. 
}
 {Our extensive spectroscopic observations of the classical nova 
V339~Del allowed us to map its evolution from the very early 
phase after its explosion. It is evident that the nova was not 
evolving according to the current theoretical prediction. 
The unusual non-spherically symmetric ejecta of nova V339~Del and 
its extreme physical conditions and evolution during and after the 
fireball stage represent interesting new challenges for 
the theoretical modelling of the nova phenomenon. 
}
\maketitle
%
%
\section{Introduction}

According to the first report on Nova Delphini 2013 
(PNV J20233073+2046041, V339~Del), announced by the Central 
Bureau Electronic Telegram No.~3628, the nova was 
discovered by Koichi Itagaki on 2013 Aug. 14.584 UT at the 
unfiltered brightness of 6.8 mag, and its progenitor was 
identified by Denisenko et al. as the blue star 
USNO-B1.0 1107-0509795 ($B\sim 17.2-17.4$, $R\sim 17.4-17.7$). 
First optical spectra were obtained on 2013 Aug.~14.844 by observers 
participating in the {\it Astronomical Ring for Access 
to Spectroscopy (ARAS)} project\footnote{
http://www.astrosurf.com/aras/Aras$\_ $DataBase/Novae/Nova-Del-2013.htm} 
\citep[][]{shore+13a} and by \cite{darnley+13} on 2013 Aug. 14.909. 

A first description of the multicolour optical photometry was 
provided by \cite{munari+13a}. They estimated the peak brightness 
of V339~Del to $V\sim 4.43$, reached at Aug. 16.44 UT, that is, 
$\sim 1.85$ days after its discovery. 
An increase of the observed colour index $B-V$ from $\sim 0.1$ 
to $\sim 0.55$ between Aug. 15.0 and 19.2 
\citep[see Fig.~2 of][]{munari+13a} corresponded to a change of 
the spectral type of the expanding envelope from $\sim$A2 to $\sim$F8 
\citep[cf.][]{cox00}. The following gradual decrease of the 
$B-V$ index after Aug.~20 suggested a dramatic change of the optical 
spectrum of the nova.

Spectroscopic observations during this early stage of the 
nova evolution were consistent with its photometric behaviour. 
\cite{tar+shak} reported that the spectral energy distribution 
(SED) of their pre-maximum spectrum, taken on Aug. 15.8, was 
similar to a late-A or early-F-type spectrum. The line spectrum, 
which was dominated by P~Cyg-type hydrogen lines with a pronounced 
absorption component 
\citep[see][]{darnley+13,shore+13a,tomov+13,tarasova,munari+13b}, 
also suggested that the effective temperature of the envelope 
probably is $\lessapprox 10^4$\,K. 
During Aug. 19, \cite{darn+bode} observed a significant weakening 
of the \ion{H}{i} absorption components, while those of the 
\ion{Fe}{ii}, \ion{He}{i}, and \ion{O}{i} profiles were still clearly 
present. This evolution indicated a distinctive cooling of the 
expanding envelope. In the continuum, the temperature decrease 
was manifested by a flattening of the optical SED almost without 
traces of the Balmer jump, as observed by \cite{tar+shak} on 
Aug. 19.9. 
This behaviour signalized the end of the early optically thick 
phase \citep[also called the {\em fireball} or iron-curtain 
stage, see][]{shore08}, when the expanding shell transfers 
the inner energetic photons to its optically thick/thin 
interface. When the nova shell reaches a maximum radius, 
the optical depth of its outer parts starts to decrease, and 
the WD pseudophotosphere progressively shrinks and becomes 
hotter. As a result, the spectrum significantly changes in both 
the continuum and lines, shifting the maximum of its SED 
to shorter wavelengths. 
For nova V339~Del this transition occurred during 
Aug. 20--22, as described by \cite{shore+13b}. 

In this paper we describe the evolution of the fundamental 
parameters, $L_{\rm WD}$, $R_{\rm WD}$, and $T_{\rm eff}$ of 
nova V339~Del from its detection to the onset of a stage with 
a harder spectrum (around day 40, when the first significant 
X-ray flux occurred). 
We note that the Raman-scattered 6825\,\AA\ line was observed for
the first time in the spectrum of a classical nova, while it is
commonly identified only for symbiotic systems. The variation 
of the \ha\ and Raman 6825\,\AA\ line profiles allowed 
us to determine the ionization structure of the ejecta during 
this early period of the nova evolution. 
Section~3 presents the results, which are interpreted 
and summarized in Sects.~4 and 5. 

\section{Observations}

Spectroscopic observations of V339~Del were secured at different 
observatories and/or private stations: 

(i)
At the the Crimean Astrophysical Observatory 
with the 2.6 m Shajn telescope, 
using a {\small SPEM} spectrograph in the Nasmith focus; 
the detector was a SPEC-10 CCD camera (1340$\times$100 pixel). 

(ii)
At the Fujii Kurosaki Observatory with a 0.4 m SCT F10 (Meade) 
telescope, using a {\small FBSPEC-III} spectrograph and a CCD 
camera ML6303E(FLI) (3072$\times$2048 pixel) as the detector. 

(iii) 
At the Observatory de la Tourbi\`ere with a 0.355 m Schmidt-Cassegrain 
telescope (Celestron C14) equipped with an eShel spectrograph 
(Shelyak; optical fiber of 50\,$\mu$m) mounted at the f/d7 focus. 
The detector was an ATIK\,460EX CCD camera (pixel size of 
4,54\,$\mu$m, binning 2$\times$2 mode; pixel of 9.08\,$\mu$m). 

(iv) 
At the Santa Maria de Montmagastrell Observatory, T\`arrega 
(Lleida) Spain, using a Control Remote Telescope SC16 equipped 
with a spectrograph {\small B60050-VI}; the detector was an 
ATIK\,460EX CCD camera. 

(v)
At a private station in Idaho Falls with a 0.35 m Celestron C14 
telescope, using a LISA spectrograph from Shelyak Instruments; 
the detector was an ATIK\,460E CCD camera. 

(vi) 
At the Castanet Tolosan Observatory with a 0.2 m F4 Newton telescope, 
using the eShel model of the echelle spectrograph (Shelyak) and 
a CCD camera ATIK\,640EX. On Oct.~1.923 2013, a 0.28 m F6 SCT 
telescope was used. 

(vii) 
At the private station Fayssac with a 0.25 m telescope, using 
the {\small Alpy 600} spectrograph (Shelyak) in the Newton focus; 
the detector was an ATIK\,314L CCD camera. 

(viii)
At the Observatoire du Pilat, with a 0.25 m LX200 (Meade) telescope, 
using the {\small Alpy 600} spectrograph (Shelyak); the detector 
was SBIG ST-8300 CCD camera wit a Kodak KAF 8300 chip. 

(ix) 
At the private station Chelles, 
with a 0.35 m SCT F11 (Celestron) telescope, using the eShell 
cross-dispersed echelle spectrograph (Shelyak) and a CCD camera 
ATIK\,460EX with ICX694 (Sony) sensor. 

(ix)
At the Durtal Observatory with a 0.51 m F/5 telescope, using 
the eShel model of the echelle spectrograph (Shelyak); the detector
was a KAF-3200ME CCD camera. 

(x) 
At the Mirranook Observatory with a 0.25 m F6.4 SCT (Celestron) 
telesope, using a LISA spectrograph (Shelyak); the detector was 
an ATIK\,314L CCD camera with a 23\,$\mu$m slit. 

(xi)
At the private station of the Societe Astronomique de 
Bourgogne-Dijon with a 0.2 m telescope (Celestron) using a LISA
spectrograph (Shelyak); the detector was an ATIK\,314L CCD 
camera. 

(xii) 
At the private stations, Manhattan (Illinois) and Grand Lake 
(Colorado) by a 0.25 m LX200 SCT telescope using the 
{\small Alpy 600} spectrograph (Shelyak). The detector
was an ATIK\,314L CCD camera. 

For the purpose of modelling the SED we used low-resolution 
spectra ($R\sim 500-1000$, Table~1), while the medium-resolution 
spectra ($R\sim$\,10000, Table~2) served to analyse variations 
in the line profiles. Relative flux units were converted into 
absolute fluxes with the aid of the (near-)simultaneous 
$BVR_{\rm C}I_{\rm C}$ photometry of \cite{munari+13a} 
and/or $UBV$ photometry from the AAVSO database. 
Magnitudes were converted to fluxes according to the
calibration of \cite{hk82}. 
Observations were dereddened with $E_{\rm B-V}$ = 0.18 
\citep[][]{munari+13b,shore+13c,chochol+14} by using the 
extinction curve of \cite{c+89}. 
Then we scaled the dereddened spectra to the $V$ flux-point. 

We selected only spectra with a good match at all available 
photometric flux-points, but especially in the $B$ band. 
The measured photometric magnitudes were corrected for emission 
lines to obtain fluxes of the true continuum \citep[see][]{sk07}. 
Some spectra showed an excess or depression for 
$\lambda \la 4500$\,\AA. They were not used in our analysis, 
because the continuum profile sensitively depends on the 
effective temperature of the shell -- the primary parameter 
of our SED-fitting analysis (Sect.~3.1). 

We also used near-IR $JHKLM$ photometry of nova V339~Del 
as published by \cite{burlak+13}, \cite{gehrz+13}, 
\cite{cass+13a,cass+13b} and \cite{shenavrin+13}. 
Resulting parameters were scaled to a distance of 3\,kpc 
\citep[][]{chochol+14}. 
%
%
\begin{table}
\caption[]{Log of low-resolution spectroscopic observations}
\begin{center}
\begin{tabular}{cccrrl}
\hline
\hline
Day$^{1}$ & Julian date & Region&$T_{\rm exp}$& R$^{2}$&Observer  \\
2013 Aug. & JD~2\,4565..&  [nm] &   [s]       &        &          \\
\hline \\[-2mm]
14.905    & 19.405      & 373--742 & 623   & 754       & Guarro \\
14.934    & 19.434      & 373--743 & 623   & 771       & Guarro \\
14.972    & 19.472      & 372--743 & 528   & 755       & Guarro \\
15.004    & 19.504      & 371--743 & 423   & 741       & Guarro \\
15.483    & 19.983      & 355--950 & 300   & 500       & Fujii-san \\
15.496    & 19.996      & 380--730 & 1778  & 1306      & Bohlsen\\
15.634    & 20.134      & 355--950 & 300   & 500       & Fujii-san \\
15.804    & 20.304      & 333--758 &  10   & 1000      & Tarasova \\
15.869    & 20.369      & 371--729 & 791   & 675       & Terry \\
15.996    & 20.496      & 371--743 & 370   & 784       & Guarro \\
16.198    & 20.698      & 372--757 & 773   & 712       & Edlin \\
16.470    & 20.970      & 355--960 & 300   & 500       & Fujii-san \\
16.512    & 21.012      & 355--960 & 240   & 500       & Fujii-san \\
16.553    & 21.053      & 355--960 & 240   & 500       & Fujii-san \\
16.595    & 21.095      & 355--960 & 240   & 500       & Fujii-san \\
16.637    & 21.137      & 355--960 & 220   & 500       & Fujii-san \\
16.740    & 21.240      & 355--960 & 400   & 500       & Fujii-san \\
16.885    & 21.385      & 371--729 & 3193  & 682       & Terry \\
16.950    & 21.450      & 379--722 & 1914  & 912       & Favaro \\
17.134    & 21.634      & 375--700 & 720   & 514       & Graham \\
17.265    & 21.765      & 375--740 & 747   & 884       & Edlin \\
17.473    & 21.973      & 355--960 & 400   & 500       & Fujii-san \\
17.543    & 22.043      & 355--960 & 400   & 500       & Fujii-san \\
17.710    & 22.210      & 355--960 & 400   & 500       & Fujii-san \\
17.898    & 22.398      & 361--738 & 3854  & 513       & Antao \\
18.189    & 22.689      & 367--740 & 2331  & 758       & Edlin \\
18.456    & 22.956      & 355--960 &  280  & 500       & Fujii-san \\
18.520    & 23.020      & 355--960 &  300  & 500       & Fujii-san \\
18.585    & 23.085      & 355--960 &  220  & 500       & Fujii-san \\
18.873    & 23.373      & 365--735 & 2061  & 513       & Antao \\
18.943    & 23.443      & 365--735 & 2031  & 513       & Antao \\
19.151    & 23.651      & 375--739 &  773  & 938       & Edlin \\
19.480    & 23.980      & 355--960 &  260  & 500       & Fujii-san \\
19.550    & 24.050      & 355--960 &  200  & 500       & Fujii-san \\
19.849    & 24.349      & 368--743 &  327  & 760       & Guarro \\
19.887    & 24.387      & 333--757 &   10  & 1000      & Tarasova \\
21.691    & 26.191      & 365--929 &   91  & 500       & Fujii-san \\
21.822    & 26.322      & 368--735 & 1367  & 591       & Thizy \\
26.161    & 30.661      & 360--740 &  398  & 503       & Graham \\
28.477    & 32.977      & 355--960 &   91  & 500       & Fujii-san \\
Sept.     &             &          &       &           &           \\
13.536    & 49.036      & 355--960 &   96  & 500       & Fujii-san \\
13.748    & 49.248      & 333--757 &   10  & 1000      & Tarasova \\
20.542    & 56.042      & 355--960 &  400  & 500       & Fujii-san \\
20.864    & 56.364     &663--1020~~& 4310  & 1085      & Guarro \\
\hline
\end{tabular}
\end{center}
  $^{1}$~mid of observation in UT,
  $^{2}$~average resolution
\end{table}  
%
%
%
\begin{table}
\caption[]{Log of medium-resolution spectroscopic 
           observations$^{1)}$}
\begin{center}
\begin{tabular}{cccccl}
\hline
\hline
Day$^{2)}$ & JD   & Region$^{3)}$&$T_{\rm exp}$& 
$F_{\rm cont.}^{4)}$&Observer  \\
 Aug.   & 2\,4565..& [nm] &   [s]       &        &          \\
\hline \\[-2mm]
 14.844  & 19.344  & 418--732   &3636   & 9.7$^{5)}$  & Garde \\
 14.885  & 19.385  & 418--732   &3349   &10.0$^{5)}$  & Garde \\
14.985  & 19.485  & \ha\       &3683   &11.9   & Garde \\ 
15.027  & 19.527  & \ha\       &3624   &12.7   & Garde \\
 15.062  & 19.562  & \ha\       &2422   &13.4   & Garde \\
15.850  & 20.350  & \ha\       &3180   &32.1   & Garde \\
 15.885  & 20.385  & \ha\       &2840   &32.9   & Garde \\
 15.885  & 20.385  & Raman      &2840   &29.3   & Garde \\
 16.910  & 21.410  & \ha\       &3830   &55.5   & Garde \\
 16.910  & 21.410  & Raman      &3830   &49.6   & Garde \\
 17.842  & 22.342  & \ha\       &4263   &49.0   & Garde \\
 17.967  & 22.467  & Raman      &4263   &47.4   & WR13x$^{6)}$ \\
18.044  & 22.544  & \ha\       &3074   &48.9   & Garde \\
 18.913  & 23.413  & \ha\       & 622   &48.7   & Lemoult \\
 19.050  & 23.550  & Raman      & 900   &45.2   & WR13x \\
19.050  & 23.550  & \ha\       & 900   &48.7   & WR13x \\
 19.870  & 24.370  & \ha\       &2877   &49.0   & Lemoult \\
 19.870  & 24.370  & Raman      &2877   &47.0   & Lemoult \\
20.051  & 24.551  & \ha\       & 438   &49.0    & WR13x  \\
 20.840  & 25.340  & Raman      &3659   &34.7   & Garde \\
 21.876  & 26.376  & Raman      &4027   &25.5   & Buil \\
 22.854  & 27.354  & Raman      &4575   &21.5   & Garde \\
 23.845  & 28.345  & Raman      &4921   &19.0   & Garde \\
 24.979  & 29.479  & Raman      &3618   &17.1   & Buil \\
 26.828  & 31.328  & Raman      &6040   &13.1   & Garde \\
28.849  & 33.349  & \ha\       &4546   &13.9   & Buil \\ 
 Sept.   &         &            &       &       &       \\
 01.881  & 37.381  & Raman      &2565   &7.81   & Charbonnel \\
 05.955  & 41.455  & Raman      & 827   &6.02   & WR13x  \\
 10.840  & 46.340  & Raman      &2350   &5.34   & Charbonnel \\
 21.825  & 57.325  & Raman      &6060   &3.74   & Garde \\
 22.858  & 58.358  & Raman      &2458   &3.45   & Buil \\
 Oct.    &         &            &       &       &       \\
 01.923  & 67.423  & Raman      &4555   &1.10   & Buil \\
\hline
\end{tabular}
\end{center} 
  $^{1})$~the average resolving power was 10000--11000, 
  $^{2})$~mid of observation in UT,
  $^{3})$~the used wavelength range, 
  $^{4})$~the continuum flux at the \ha\ 
          or the Raman 6825\,\AA\ line in $10^{-12}$\ecsa, 
  $^{5)}$~$F_{\rm cont.}$ at the \ha\ line, 
  $^{6})$~WR13x-collaboration team 
\end{table}
%
%
%
\begin{figure*}
\begin{center}
\resizebox{18cm}{!}{\includegraphics[angle=-90]{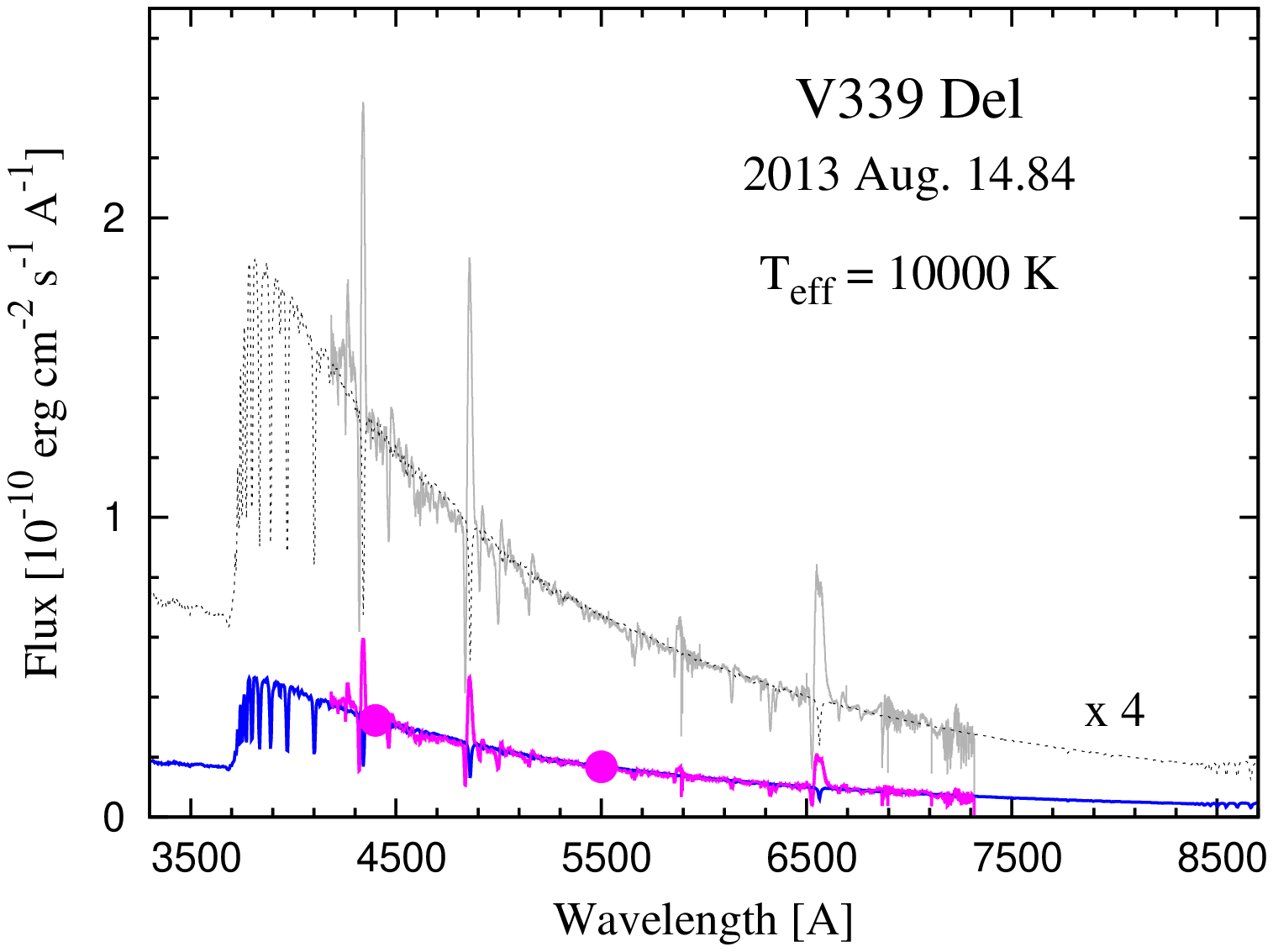}
                      \includegraphics[angle=-90]{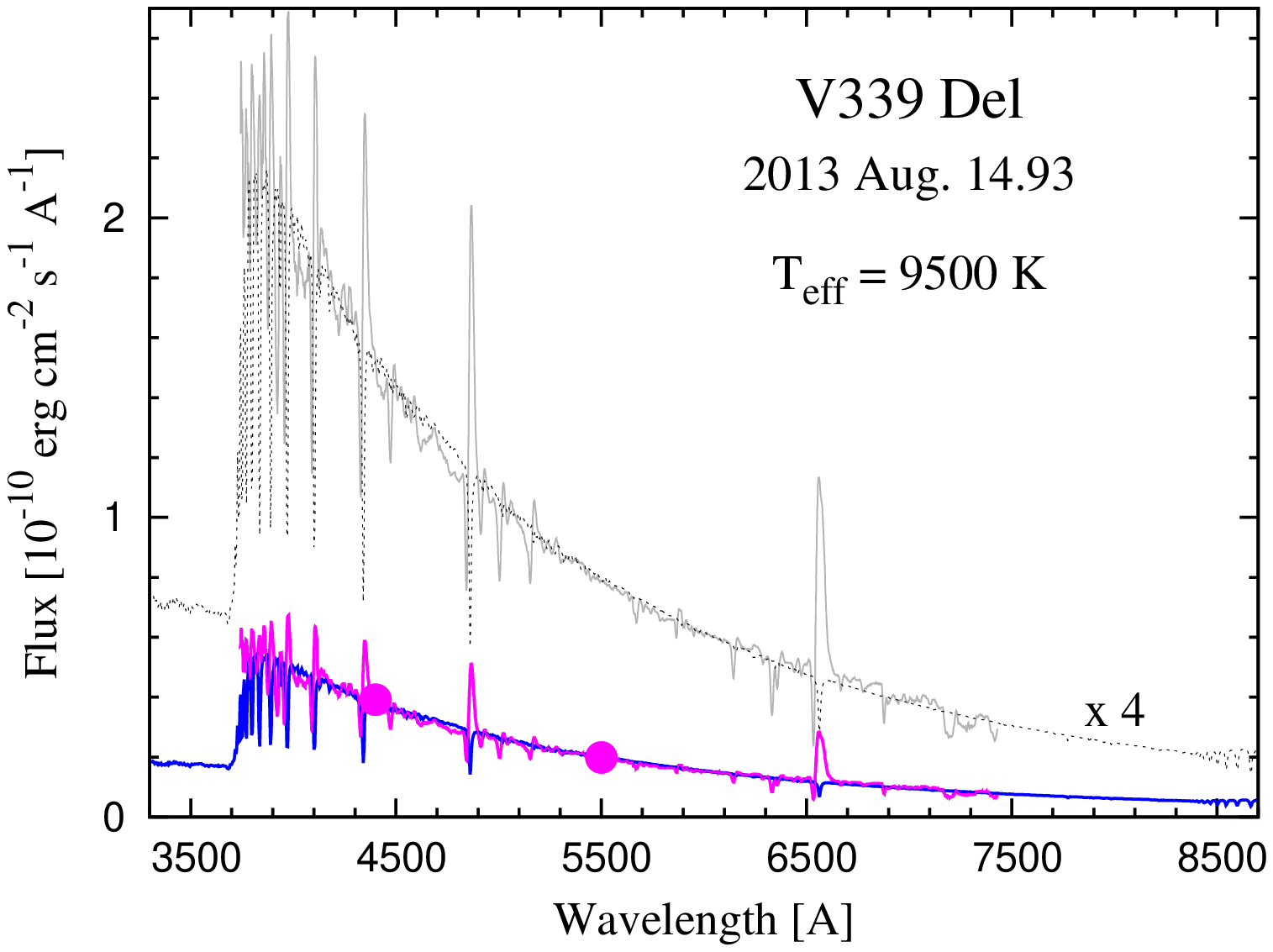}
                      \includegraphics[angle=-90]{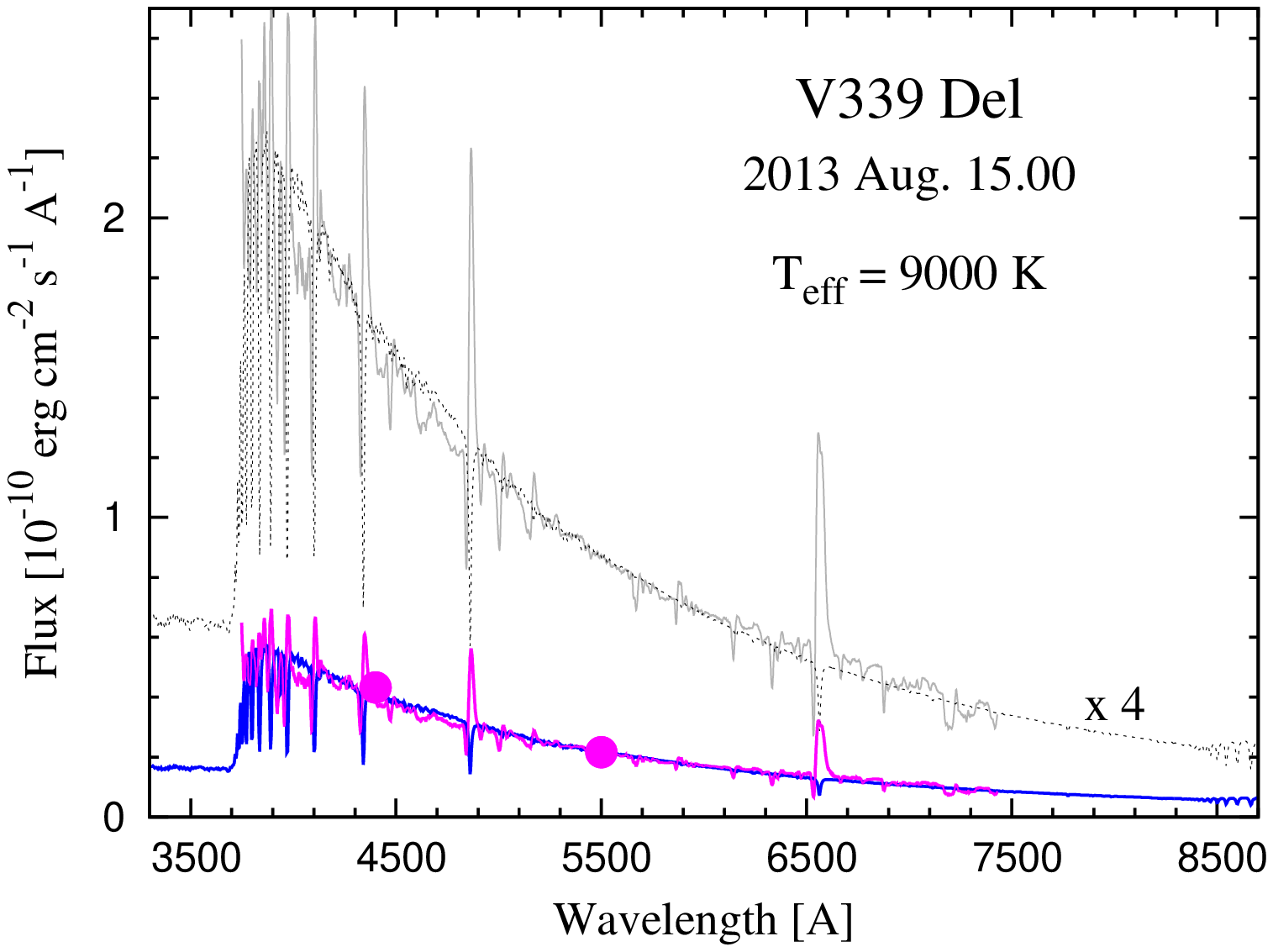}}
\resizebox{18cm}{!}{\includegraphics[angle=-90]{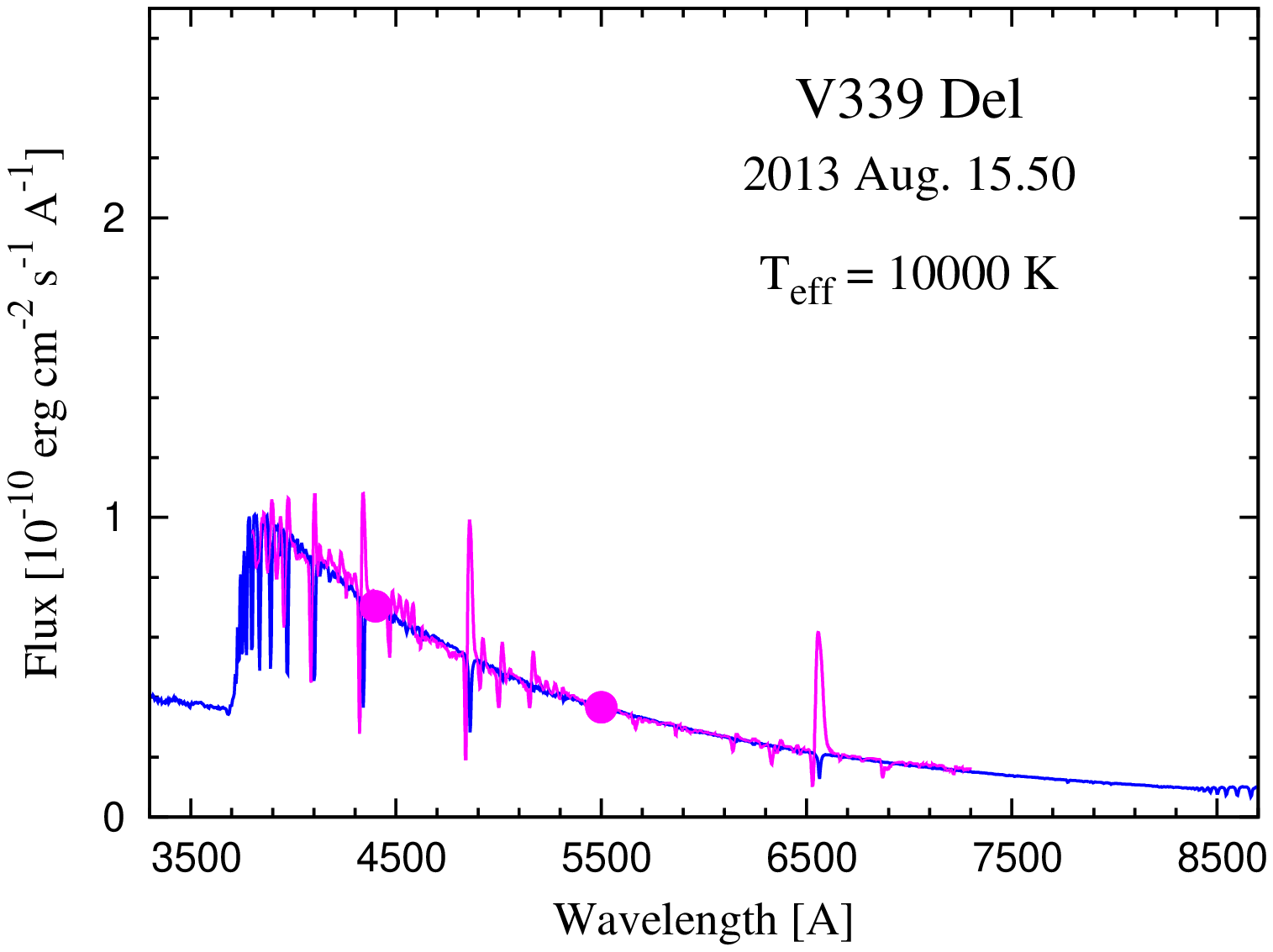}
                      \includegraphics[angle=-90]{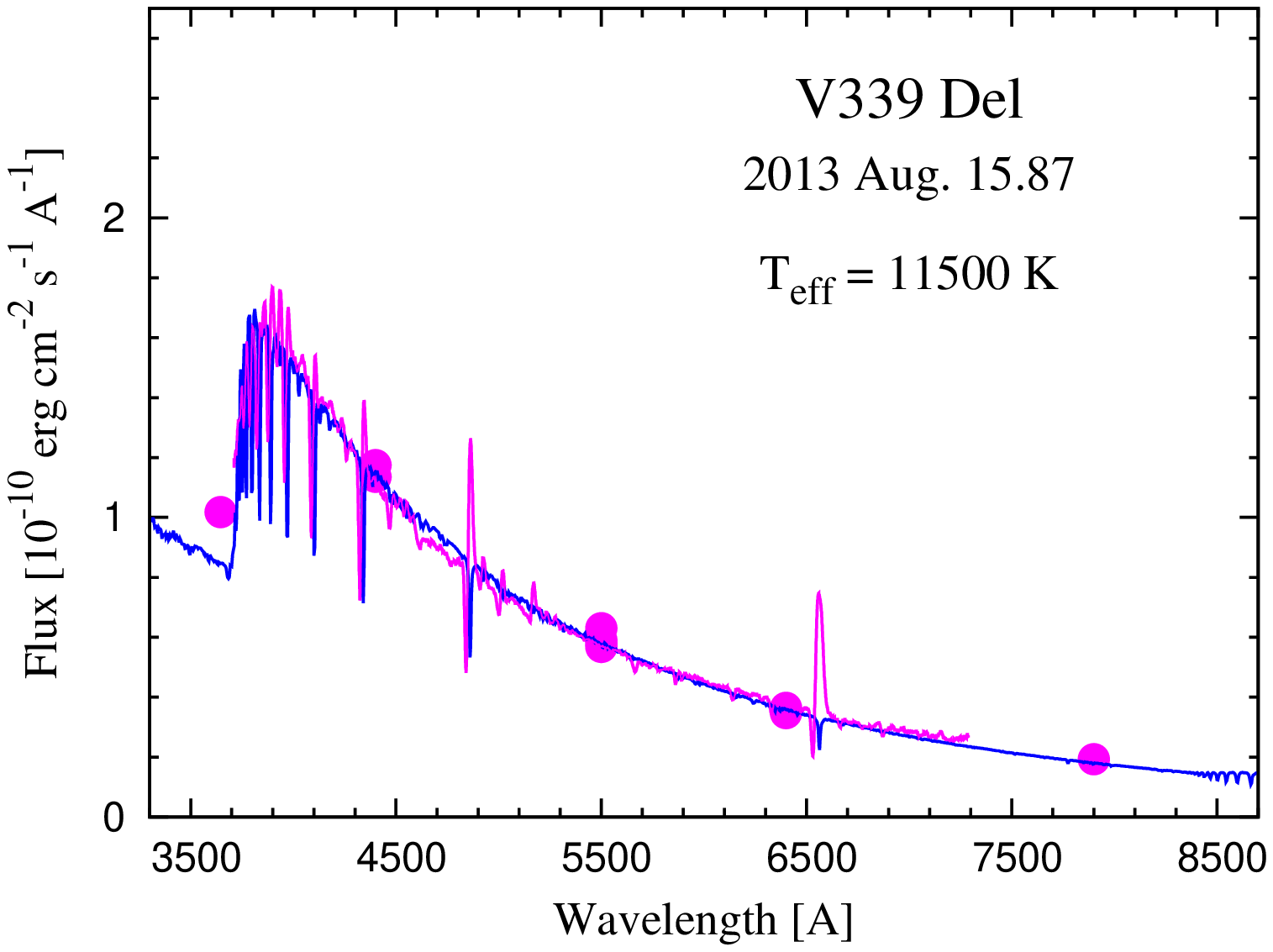}
                      \includegraphics[angle=-90]{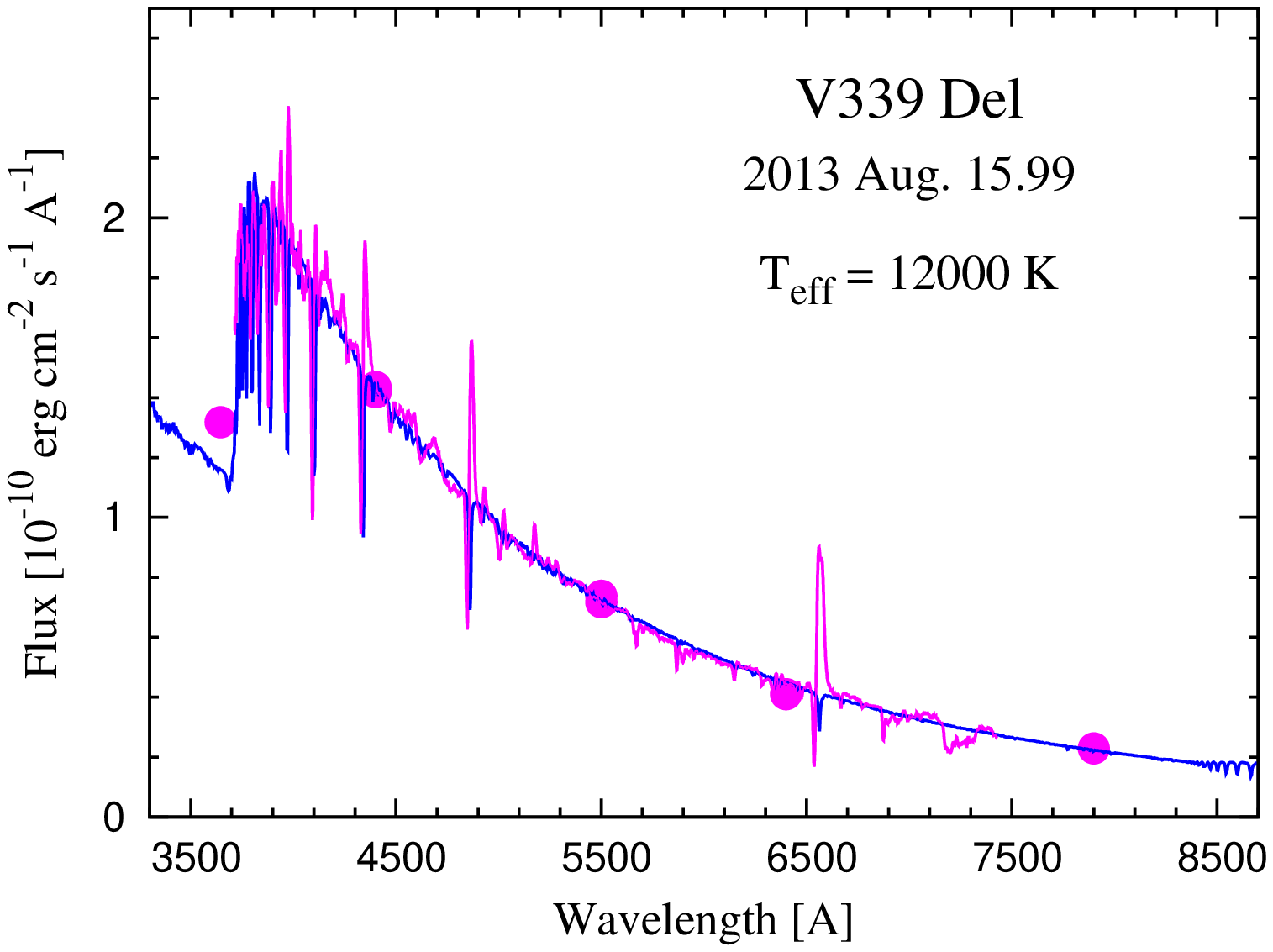}}
\resizebox{18cm}{!}{\includegraphics[angle=-90]{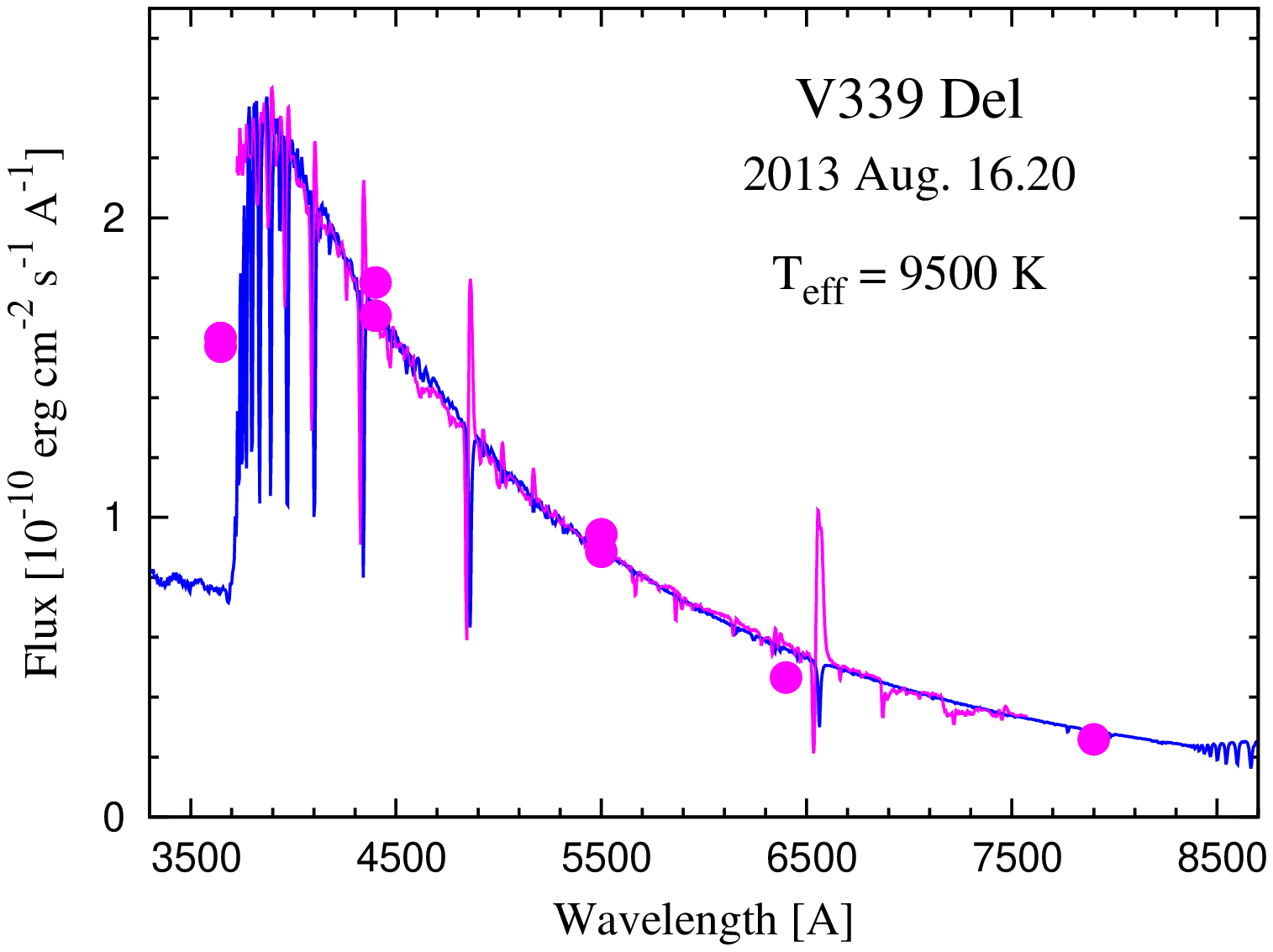}
                      \includegraphics[angle=-90]{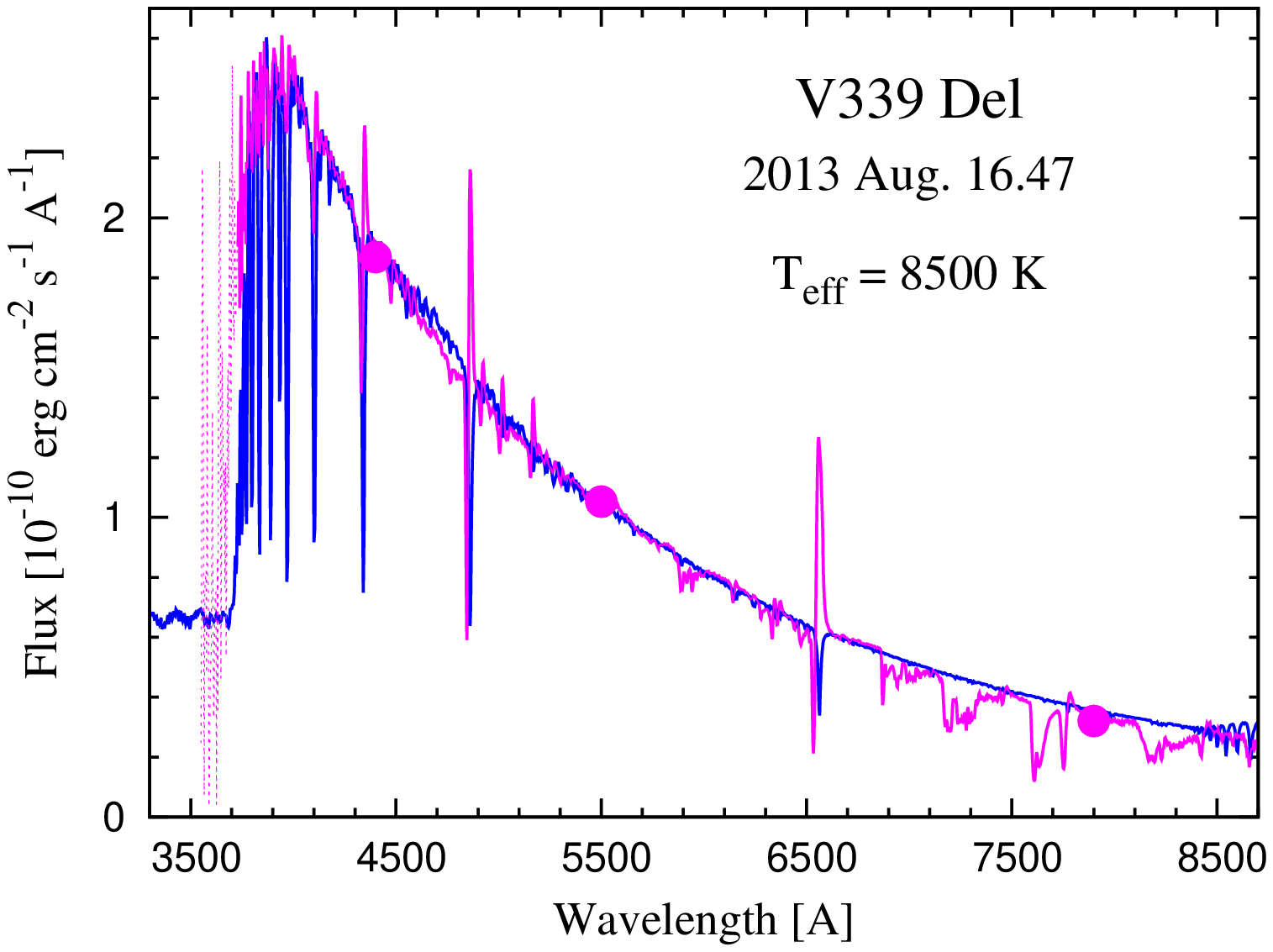}
                      \includegraphics[angle=-90]{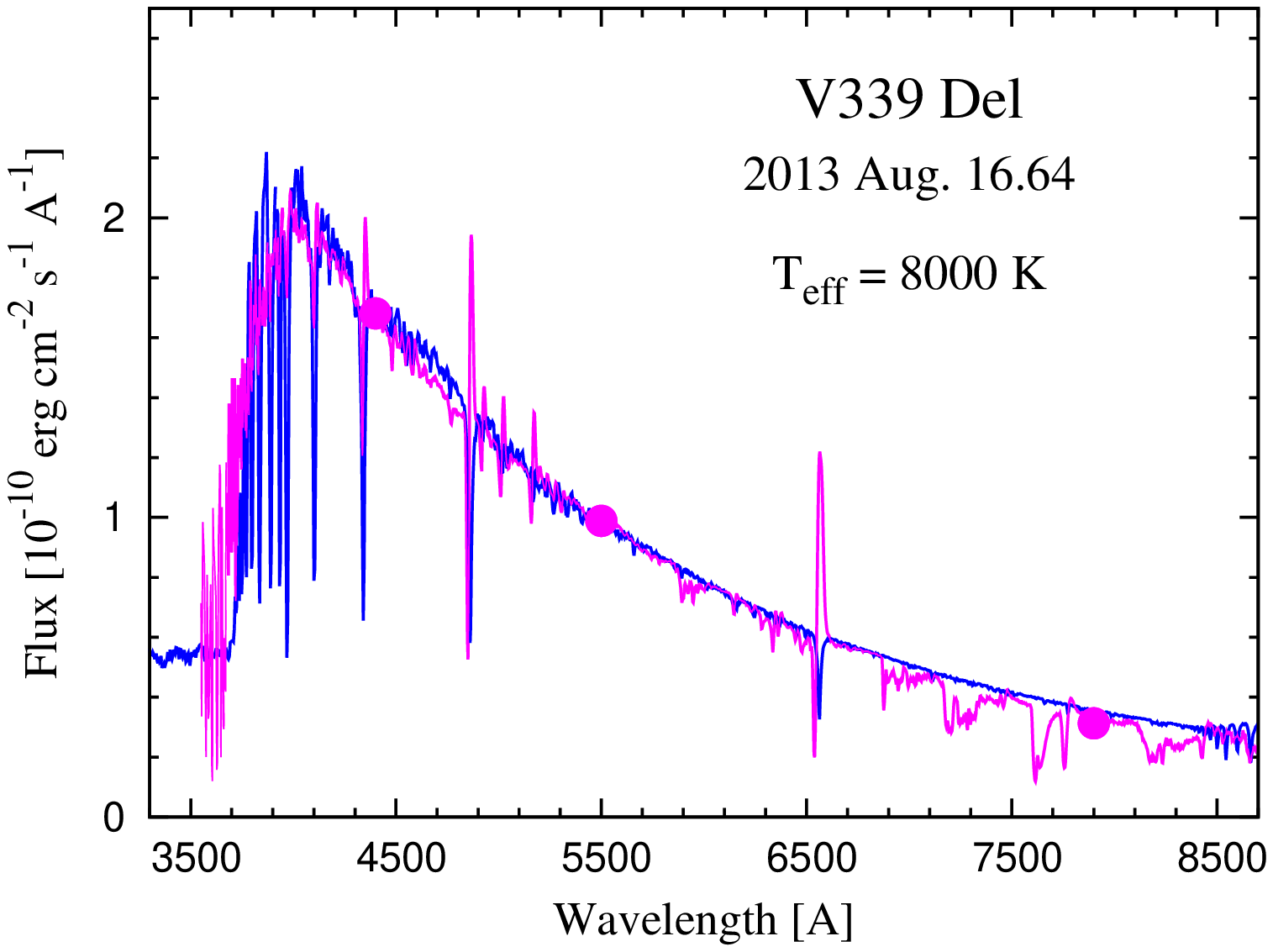}}
\resizebox{18cm}{!}{\includegraphics[angle=-90]{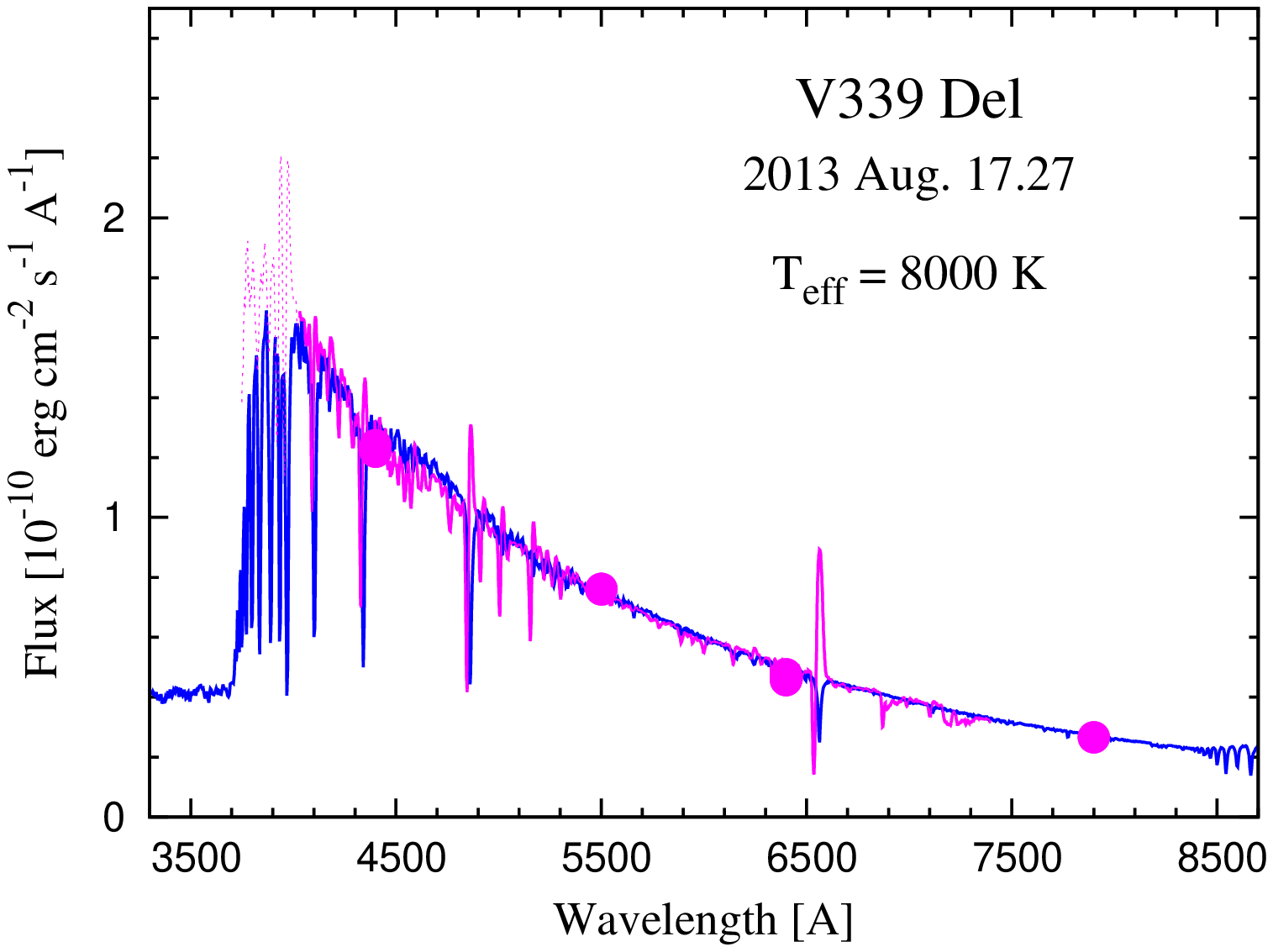}
                      \includegraphics[angle=-90]{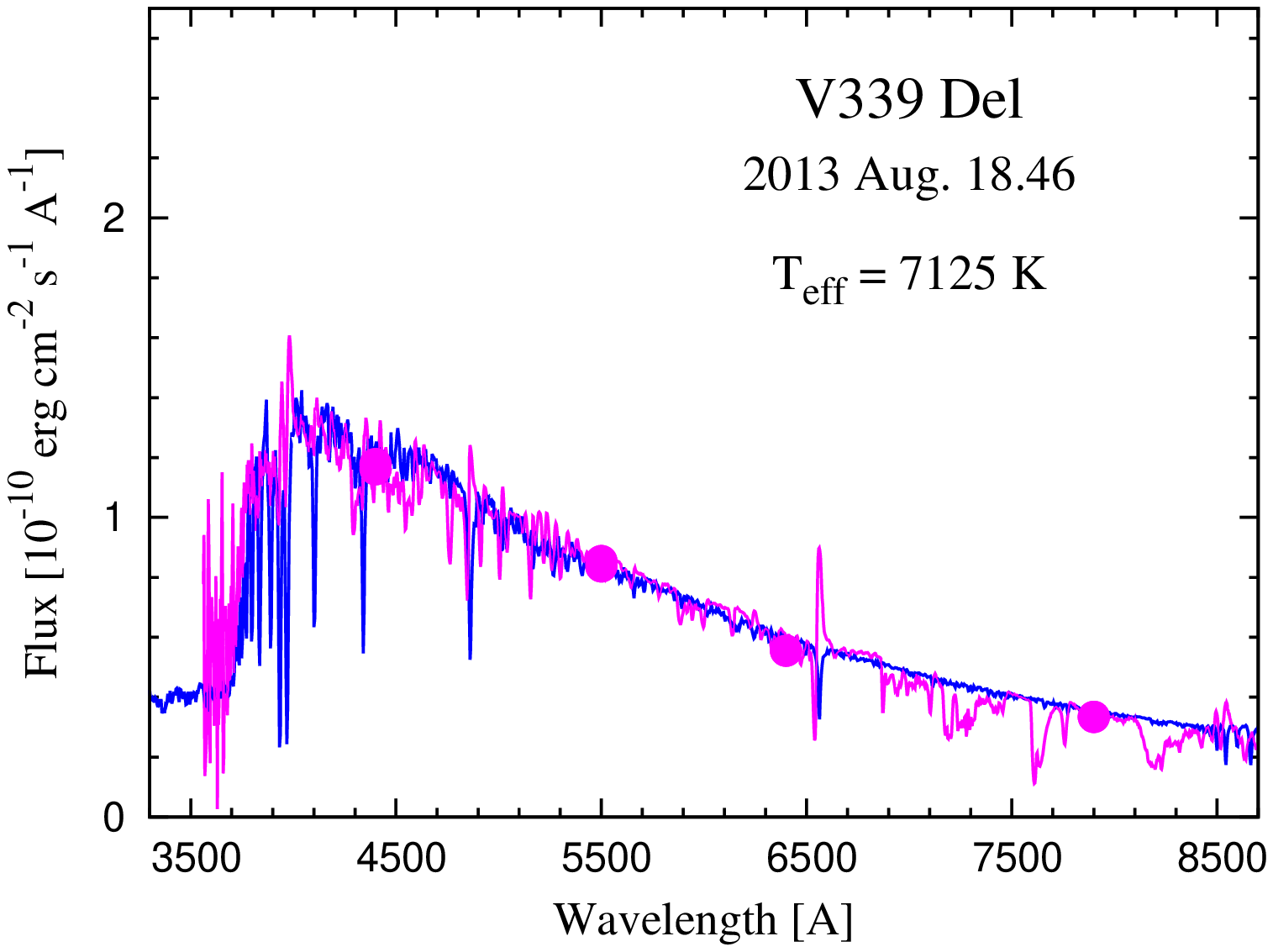}
                      \includegraphics[angle=-90]{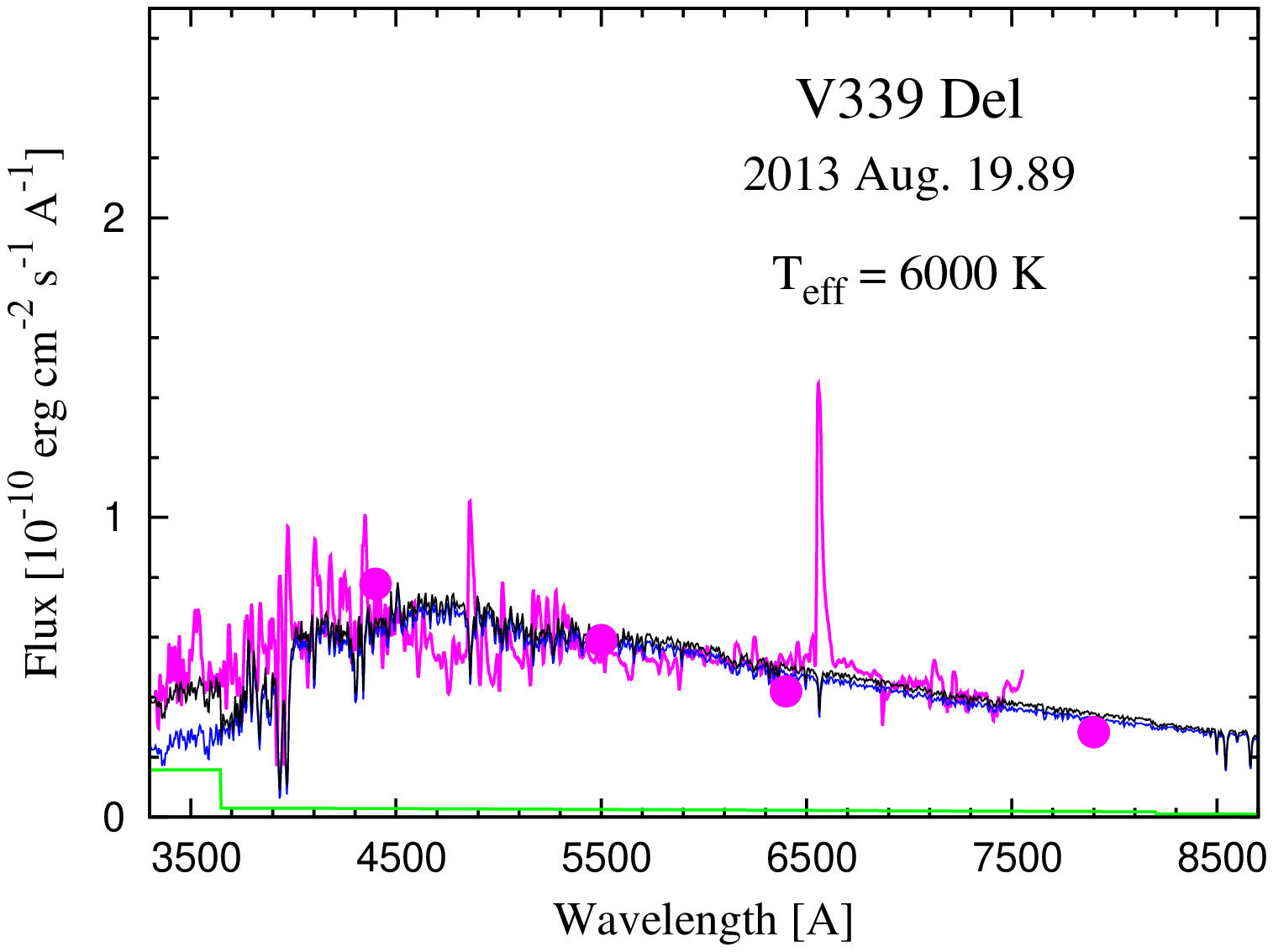}}
\resizebox{18cm}{!}{\includegraphics[angle=-90]{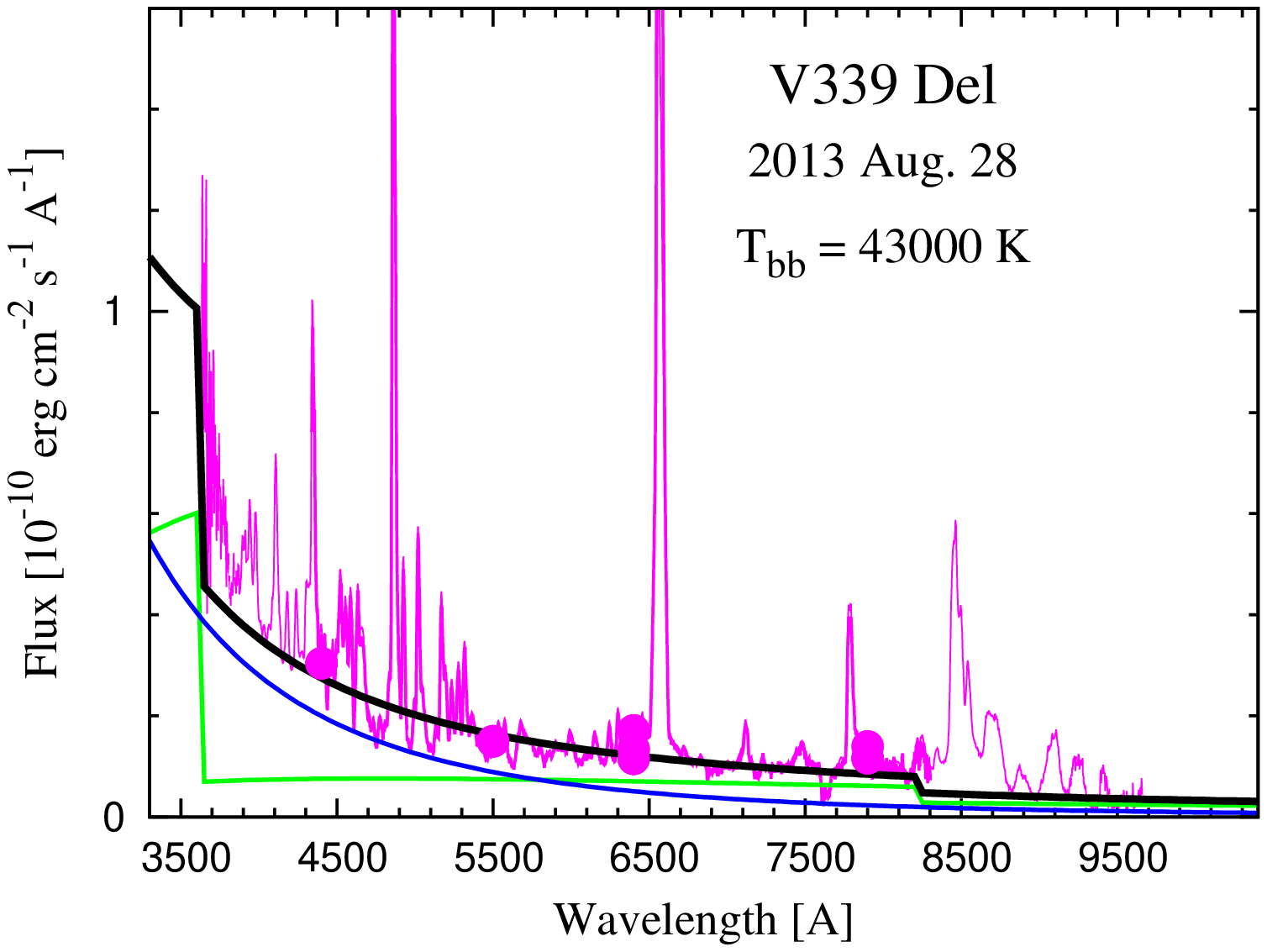}
                      \includegraphics[angle=-90]{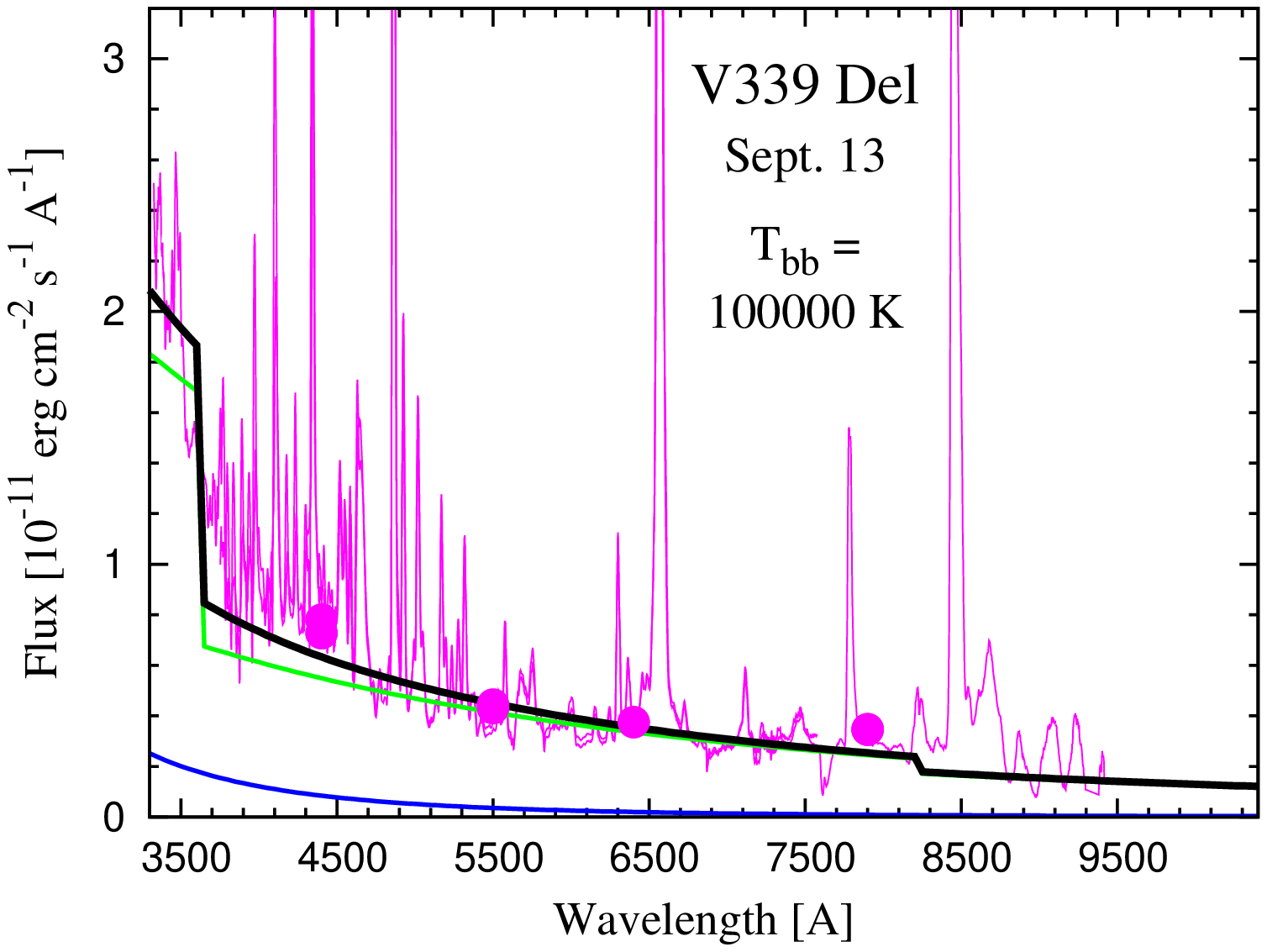}
                      \includegraphics[angle=-90]{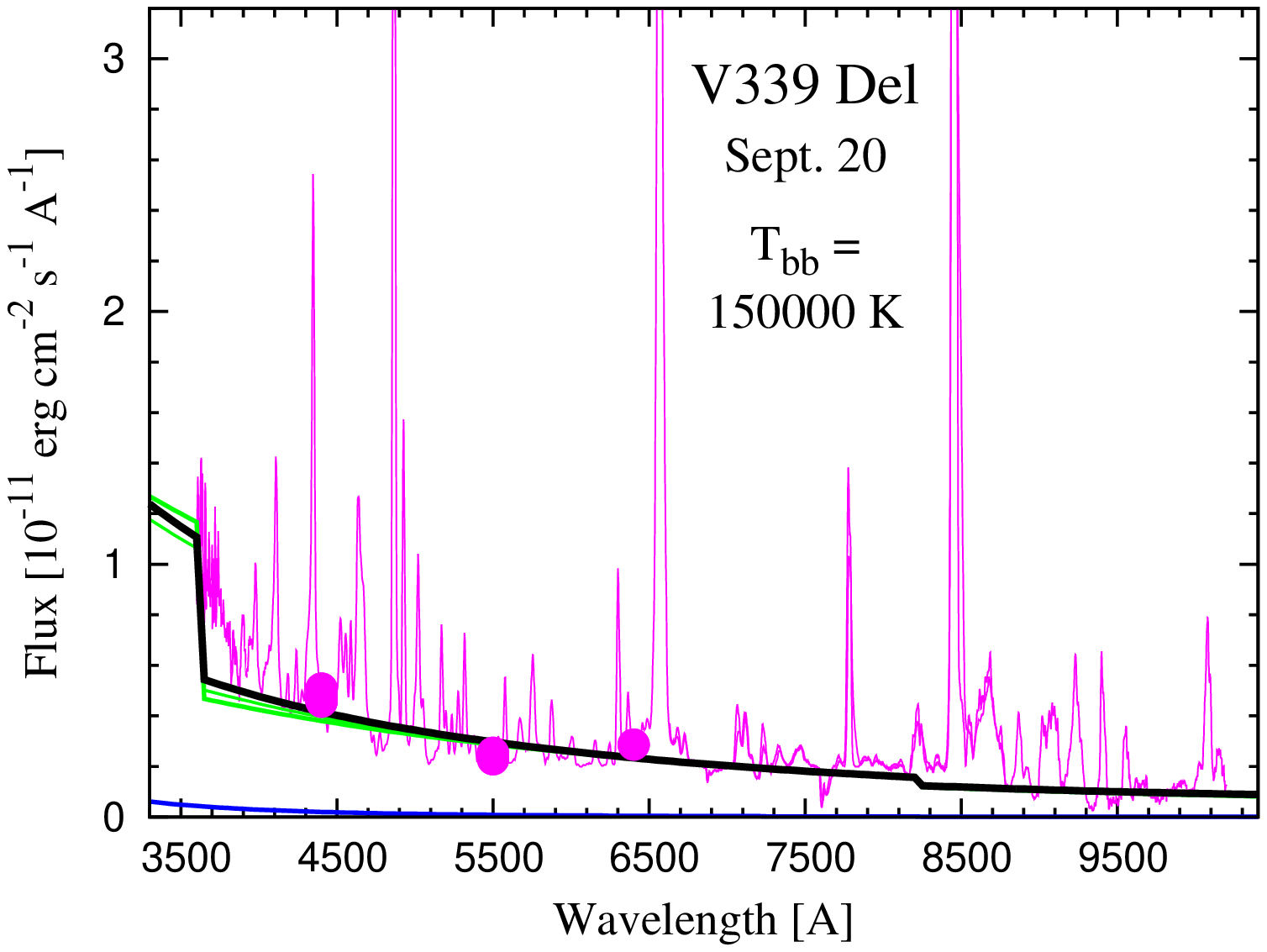}}
\end{center}
\caption[]{
Comparison of the observed (in magenta: spectrum and photometric 
flux-points) and model SEDs of V339~Del. During the fireball 
stage (2013, Aug.~14.84 to 19.89), the model SED is represented 
by a synthetic spectrum (blue line, Eq.~(\ref{eq:fl1})), 
while during the following stage with a harder spectrum 
(bottom row of panels) the model SED (black line) is given 
by a superposition of the radiation from the WD pseudophotosphere 
(blue line) and the nebula (green line) according to 
Eq.~(\ref{eq:fl2}).
          }
\label{fig:sedopt}
\end{figure*}
%
%
\begin{figure}
\begin{center}
\resizebox{7.1cm}{!}{\includegraphics[angle=-90]{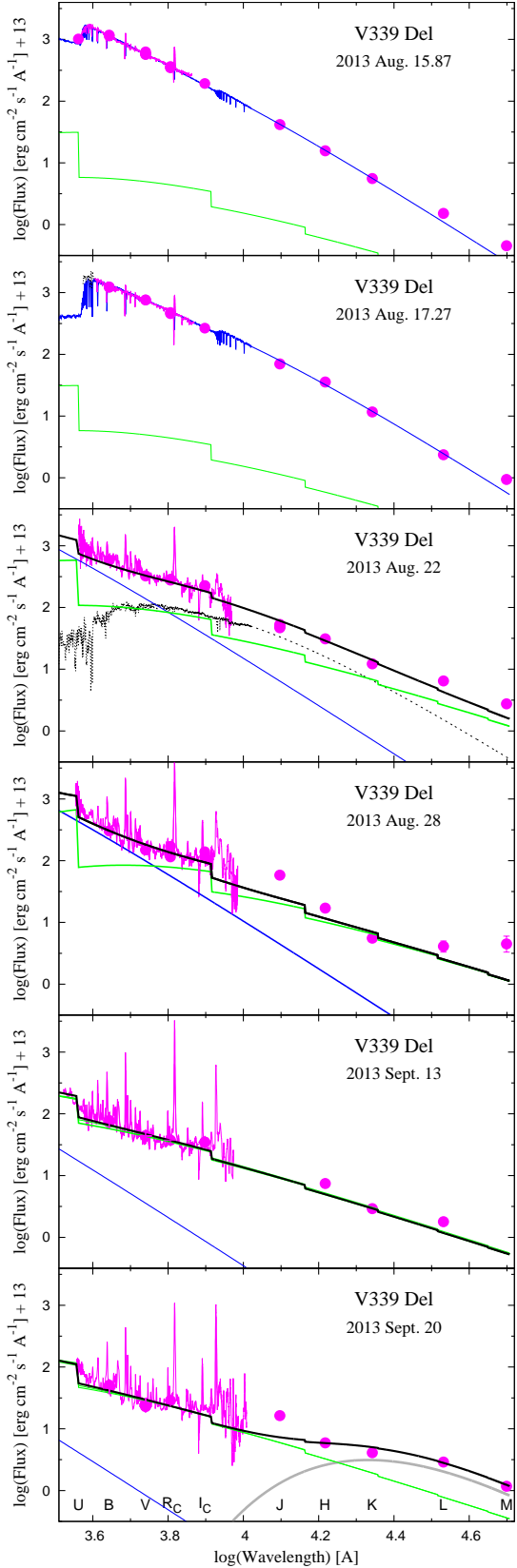}}
\end{center}
\caption[]{
Optical/near-IR SEDs of nova V339~Del during the fireball 
stage and beyond. Denotation of lines 
and points is the same as in Fig.~\ref{fig:sedopt}. 
In addition, the dotted line in the third panel from the top 
represents a remnant of the WD shell and the grey line in 
the bottom panel denotes the dust component (Sect.~3.2.2). 
The nebular continuum during the fireball stage 
(top two panels) was only roughly estimated from 
the \ha\ flux for comparison. 
          }
\label{fig:sedir}
\end{figure}

\section{Analysis and results}

The primary aim of this section is to match the observed SED in 
our spectra by an appropriate model, and in this way determine 
the physical parameters of the emitting region at a given time 
of the nova evolution. 

\subsection{Model SEDs}

During the fireball stage of novae, the radiation of their 
pseudophotospheres resembles that produced by a star of spectral 
type A to F \citep[e.g.][ Sect.~1 here]{warner08}. 
Accordingly, we compared the observed spectrum, $F(\lambda)$, 
with a synthetic one, $\mathcal{F}_{\lambda}(T_{\rm eff})$, 
i.e. 
%
\begin{equation}
 F(\lambda) = \theta_{\rm WD}^2\,
              \mathcal{F}_{\lambda}(T_{\rm eff}),
\label{eq:fl1}
\end{equation}
where $\theta_{\rm WD} = R_{\rm WD}/d$ is the angular radius 
of the WD pseudophotosphere and $T_{\rm eff}$ its 
effective temperature. 
Fitting parameters are $\theta_{\rm WD}$ and $T_{\rm eff}$, 
which define the effective radius of the shell as 
$R_{\rm WD} = \theta_{\rm WD} \times d$ and its luminosity 
$L_{\rm WD} = 4\pi d^2 \theta_{\rm WD}^2 T_{\rm eff}^4$ for 
the distance to the nova, $d$. 
A warm photospheric radiation of the nova was indicated 
already by our first spectrum from Aug.~14.844. 
The last spectrum showing unambiguously a dominant contribution 
from the warm stellar pseudophotosphere in the optical was 
observed on Aug.~19.887. Therefore, we associate this date 
with the end of the fireball stage of the nova V339~Del 
(see also Sect.~4.2.3). 

In the SED-fitting analysis we compared a grid of synthetic 
models with the observed spectrum (1) and selected the one 
corresponding to a minimum of the reduced $\chi^2$ 
function. The continuum fluxes were estimated by eye. 
In this way we obtained the model variables $\theta_{\rm WD}$ 
and $T_{\rm eff}$. The grid of atmospheric models was prepared 
from that of \cite{munari+05}, for 
$T_{\rm eff} = 5\,000 - 15\,000$\,K with the step 
$\Delta T_{\rm eff} = 250$\,K and fixed other atmospheric 
parameters. The model resolution was accommodated to that 
of our spectra \citep[see][ for more details]{sk+11}. 

After the fireball stage, during the transition of the nova 
towards higher $T_{\rm eff}$, its radiation ionizes the outer 
material, which reprocesses a fraction of it into the nebular 
emission. As a result, the nebular component of radiation starts 
to rival the stellar component from the WD pseudophotosphere 
in the optical/near-IR. Because of its higher 
$T_{\rm eff}$, we observe only its long-wavelength tail 
in the optical, which can be approximated by a blackbody 
radiating at the temperature $T_{\rm bb}$. In this case, 
the observed spectrum can be expressed as 
%
\begin{equation}
 F(\lambda) = 
           \theta_{\rm WD}^2 \pi B_{\lambda}(T_{\rm bb})
         + k_{\rm n} \times \varepsilon_{\lambda}(T_{\rm e}), 
\label{eq:fl2}
\end{equation}
where the factor $k_{\rm n} = EM/4\pi d^2$ scales the total 
volume emission coefficient $\varepsilon_{\lambda}(T_{\rm e})$ 
of the nebular continuum to observations; $EM$ is the so-called 
emission measure. The variables determining the model SED (2) 
are $\theta_{\rm WD}$, $T_{\rm bb}$, $k_{\rm n}$ and $T_{\rm e}$.
Equation~(\ref{eq:fl2}) assumes that $T_{\rm e}$ and thus 
$\varepsilon_{\lambda}(T_{\rm e})$ are constant throughout 
the nebula. For the sake of simplicity, we considered only 
the contribution of the hydrogen plasma 
\citep[see e.g.][ for more details]{sk15a}. 
The first spectrum that could be matched by a superposition 
of just these two components of radiation was observed on 
Aug.~26.16. 

Our spectra taken during the first few days after the fireball 
stage indicated a persisting contribution from the warm stellar 
pseudophotosphere. This is evident from the spectra of Aug.~20.194
(J. Edlin, ARAS), 20.463, 21.691 (M. Fujii\footnote{
http://otobs.org/FBO/fko/nova/nova$\_ $del$\_ $2013.htm}), 
and 21.822 (O. Thizy, ARAS). They showed 
a steeper short-wavelength part of the spectrum for 
$\lambda \la 5000$\,\AA\ and a flat continuum at a high level 
of a few times $10^{-11}$\ecsa\ for 
$5000 \la \lambda \la 8000$\,\AA. 
The former indicates the presence of a hotter stellar component 
of radiation, while the latter is given by superposition of 
a nebular component and that from the warm pseudophotosphere. 
Therefore, we matched the observed SED in this transition 
phase by radiation components according to Eq.~(\ref{eq:fl2})
supplemented by a synthetic spectrum 
$\mathcal{F}_{\lambda}(T_{\rm eff})$ for 
$T_{\rm eff} = 5000 - 6000$\,K (see Sect.~3.2.2). 

Modelling the SED after the fireball stage is only unambiguous 
if we use data covering a broad spectral region. We 
used the spectra of Aug.~22, 28, Sept.~13, and 20, which 
cover a more extended wavelength region and were obtained 
simultaneously with the photometric near-IR observations. 
The near-IR fluxes are especially important to recognize 
the dust contribution (see Fig.~\ref{fig:sedir}). 

\begin{figure}
\begin{center}
\resizebox{\hsize}{!}{\includegraphics[angle=-90]{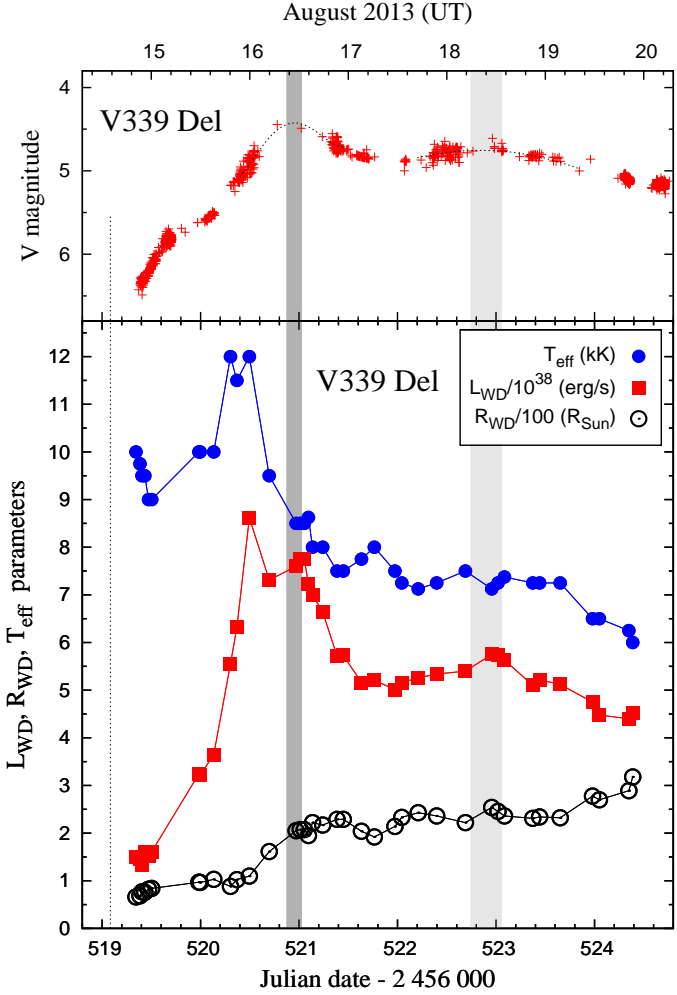}}
\end{center}
\caption[]{
Evolution of the parameters $L_{\rm WD}$, $R_{\rm WD}$, and 
$T_{\rm eff}$ of nova V339~Del during its fireball stage. 
The dotted line and grey belts mark the times of the nova 
discovery and the maxima in the $V$-light curve 
(Aug.~16.45$\pm 0.06$, 18.4$\pm 0.11$). 
Data are listed in Table~3. 
          }
\label{fig:lrt}
\end{figure}

The resulting parameters are listed in Table~3, and examples of 
corresponding models fitting the selected spectra are depicted 
in Figs.~\ref{fig:sedopt} and \ref{fig:sedir}, while 
Fig.~\ref{fig:lrt} shows the evolution of the fundamental 
parameters, $L_{\rm WD}$, $R_{\rm WD}$ and $T_{\rm eff}$, 
during the fireball stage (Aug.~14.8--19.9, 2013). 
Although that only static atmospheric models were compared, 
the resulting models express the measured continuum well. 

\subsubsection{Estimate of uncertainties}

In the range of $T_{\rm eff}$ between $\sim$\,5000 and 
$\sim$\,15000\,K, its value is sensitive to the overall profile 
of the measured continuum in the optical. This allowed us to 
estimate $T_{\rm eff}$ with a relative uncertainty of $< 4$\%. 
For the selected spectra (Sect.~2), $\Delta T_{\rm eff}$ 
was well within the temperature step in the used grid of models, 
that is, between 250 and 500\,K for $T_{\rm eff} < 10000$  and 
$> 10000$\,K. In some cases it was possible to estimate 
$\Delta T_{\rm eff}$ to $\sim$\,125\,K by interpolating the 
neighbouring models around the measured spectrum (Table~3). 

The uncertainty of the scaling factor, $\theta_{\rm WD}$, is 
primarily given by the accuracy of the measured continuum, 
which was estimated to be $< 3$\% for the used spectra. 
Different $T_{\rm eff}$ requires different scaling. 
$\Delta T_{\rm eff}$ of 250 to 500\,K corresponds to relative 
errors in $\theta_{\rm WD}$ of 2\% to 4\%. 
Thus, the largest relative errors in $\theta_{\rm WD}$ range 
between 5 and 7\%. 

The uncertainty in $L_{\rm WD}$ was determined as the root mean 
square error using the total differential of the function 
$L_{\rm WD} = 4\pi d^2 \theta_{\rm WD}^2 T_{\rm eff}^4$ 
for given uncertainties in $T_{\rm eff}$ and $\theta_{\rm WD}$. 
The relative error of $L_{\rm WD}$ was found to be 
around 10\%, depending mainly on the uncertainty in 
$T_{\rm eff}$ (see Table~3). 
%
%
\begin{table} 
\begin{center}
\caption{Physical parameters of nova V339~Del. 
        }
\begin{tabular}{crrrc}
\hline
\hline
Date                   &
$T_{\rm eff}$          &
$\theta_{\rm WD}$      &
$R_{\rm WD}$           &
$L_{\rm WD}$           \\
Aug.                   &
[K]                    &
[10$^{-10}$]           &
[$R_{\sun}$]           &
[10$^{38}$\es]         \\
\hline \\[-2mm]
\multicolumn{5}{c}{The fireball stage} \\
\hline \\[-2mm]
14.844 &10000$\pm 400$ & 5.0$\pm 0.2$ & 66$\pm  2$ &1.5$\pm 0.2$ \\
14.885 & 9750$\pm 250$ & 5.2$\pm 0.2$ & 68$\pm  2$ &1.5$\pm 0.1$ \\
14.905 & 9500$\pm 250$ & 5.8$\pm 0.2$ & 77$\pm  2$ &1.3$\pm 0.1$ \\
14.934 & 9500$\pm 250$ & 5.7$\pm 0.2$ & 76$\pm  2$ &1.6$\pm 0.1$ \\
14.972 & 9000$\pm 250$ & 6.2$\pm 0.2$ & 82$\pm  3$ &1.5$\pm 0.2$ \\
15.004 & 9000$\pm 250$ & 6.3$\pm 0.2$ & 84$\pm  3$ &1.6$\pm 0.2$ \\
15.483 &10000$\pm 250$ & 7.3$\pm 0.4$ & 97$\pm  5$ &3.2$\pm 0.4$ \\
15.496 &10000$\pm 125$ & 7.3$\pm 0.3$ & 96$\pm  4$ &3.2$\pm 0.2$ \\
15.634 &10000$\pm 250$ & 7.7$\pm 0.4$ &103$\pm  5$ &3.7$\pm 0.4$ \\
15.804 &12000$\pm 500$ & 6.6$\pm 0.3$ & 88$\pm  4$ &5.6$\pm 0.7$ \\
15.869 &11500$\pm 250$ & 7.7$\pm 0.3$ &102$\pm  4$ &6.3$\pm 0.5$ \\
15.996 &12000$\pm 500$ & 8.3$\pm 0.3$ &110$\pm  4$ &8.6$\pm 0.8$ \\
16.198 & 9500$\pm 125$ &12.1$\pm 0.5$ &161$\pm  7$ &7.3$\pm 0.5$ \\
16.470 & 8500$\pm 125$ &15.5$\pm 0.5$ &205$\pm  7$ &7.6$\pm 0.5$ \\
16.512 & 8500$\pm 250$ &15.6$\pm 0.7$ &207$\pm  9$ &7.8$\pm 0.7$ \\
16.553 & 8500$\pm 250$ &15.6$\pm 0.7$ &207$\pm  9$ &7.8$\pm 0.7$ \\
16.595 & 8625$\pm 125$ &14.6$\pm 0.5$ &195$\pm  7$ &7.2$\pm 0.4$ \\
16.637 & 8000$\pm 250$ &16.7$\pm 0.8$ &222$\pm 11$ &7.0$\pm 0.7$ \\
16.740 & 8000$\pm 250$ &16.3$\pm 0.8$ &217$\pm 11$ &6.6$\pm 0.7$ \\
16.885 & 7500$\pm 250$ &17.2$\pm 0.8$ &229$\pm 11$ &5.7$\pm 0.7$ \\
16.950 & 7500$\pm 250$ &17.2$\pm 0.8$ &229$\pm 11$ &5.7$\pm 0.7$ \\
17.134 & 7750$\pm 250$ &15.3$\pm 0.7$ &204$\pm  9$ &5.2$\pm 0.6$ \\
17.265 & 8000$\pm 125$ &14.4$\pm 0.4$ &192$\pm  5$ &5.2$\pm 0.4$ \\
17.473 & 7500$\pm 250$ &16.1$\pm 0.7$ &214$\pm  9$ &5.0$\pm 0.5$ \\
17.543 & 7250$\pm 250$ &17.5$\pm 0.8$ &233$\pm 11$ &5.2$\pm 0.6$ \\
17.710 & 7125$\pm 125$ &18.3$\pm 0.7$ &243$\pm  9$ &5.3$\pm 0.3$ \\
17.898 & 7250$\pm 250$ &17.8$\pm 0.8$ &236$\pm 11$ &5.3$\pm 0.6$ \\
18.189 & 7500$\pm 250$ &16.7$\pm 0.7$ &222$\pm  9$ &5.4$\pm 0.5$ \\
18.456 & 7125$\pm 125$ &19.1$\pm 0.7$ &254$\pm  9$ &5.8$\pm 0.3$ \\
18.520 & 7250$\pm 250$ &18.5$\pm 0.8$ &245$\pm 11$ &5.7$\pm 0.6$ \\
18.585 & 7375$\pm 125$ &17.7$\pm 0.7$ &236$\pm  9$ &5.6$\pm 0.3$ \\
18.873 & 7250$\pm 250$ &17.4$\pm 0.9$ &231$\pm 12$ &5.1$\pm 0.6$ \\
18.943 & 7250$\pm 250$ &17.6$\pm 0.9$ &234$\pm 12$ &5.2$\pm 0.6$ \\
19.151 & 7250$\pm 250$ &17.4$\pm 0.9$ &232$\pm 12$ &5.1$\pm 0.6$ \\
19.480 & 6500$\pm 250$ &20.9$\pm 0.9$ &278$\pm 12$ &4.8$\pm 0.6$ \\
19.550 & 6500$\pm 250$ &20.3$\pm 0.9$ &270$\pm 12$ &4.5$\pm 0.6$ \\
19.849 & 6250$\pm 250$ &21.7$\pm 1.0$ &289$\pm 13$ &4.4$\pm 0.7$ \\
19.887 & 6000$\pm 250$ &23.9$\pm 1.2$ &318$\pm 16$ &4.5$\pm 0.8$ \\
\hline \\[-2mm]
\multicolumn{5}{c}{Transition to a harder spectrum}              \\
\hline \\[-2mm]
21.691$^{1})$ & $>$37000 & $<$1.4     &  $<$19     & $>$22       \\
28.477$^{2})$ & $>$43000 & $<$1.0     &  $<$13     & $>$21       \\
Sept.         &          &            &            &             \\
13.642$^{3})$ & $>$100000& $<$0.12    &  $<$1.6    & $>$8.8      \\
20.703$^{4})$ & $>$150000& $<$0.046   &  $<$0.6    & $>$6.5      \\
\hline
\end{tabular}
\end{center}
Nebular component: \\
$^{1})$~ $T_{\rm e} = (2.0\pm 0.5)\times 10^4$\,K, 
         \textsl{EM} = $(2.0\pm 0.2)\times 10^{62}$\cmt \\
$^{2})$~ $T_{\rm e} = (1.5\pm 0.2)\times 10^4$\,K, 
         \textsl{EM} = $(1.8\pm 0.1)\times 10^{62}$\cmt \\
$^{3})$~ $T_{\rm e} = (4\pm 0.5)\times 10^4$\,K,                
         $\textsl{EM} = (1.4\pm 0.1)\times 10^{62}$\cmt \\
$^{4})$~ $T_{\rm e} = (5\pm 0.5)\times 10^4$\,K,
         $\textsl{EM} = (1.0\pm 0.1)\times 10^{62}$\cmt, \\ 
         dust component: $T_{\rm dust} = (1350\pm 50)$\,K, 
                    $L_{\rm dust} = (1.1\pm 0.2)\times 10^{37}$\es\
\normalsize
\end{table}
%
%

\subsection{Evolution of fundamental parameters}

\subsubsection{Fireball stage}

Following Fig.~\ref{fig:lrt}, we describe the temporal 
evolution of $T_{\rm eff}$, $R_{\rm WD}$, and $L_{\rm WD}$ of 
the WD pseudophotosphere (also called the envelope or shell) 
during its fireball stage as follows: 

Already a few hours after the discovery, the radiative output 
of the nova dominated the optical. A comparison of synthetic 
models with the observed spectra (Eq.~(\ref{eq:fl1}), 
Figs.~\ref{fig:sedopt} and \ref{fig:lrt}) showed that the nova 
envelope was radiating at $T_{\rm eff}$ = 6000--12000\,K 
during its fireball stage (Table~3). 
Our first spectra from Aug.~14.8--15.0, taken around 1.5\,days 
prior to the photometric $V$-band maximum, suggested a transient 
rapid decrease in $T_{\rm eff}$ from $\sim$\,10000\,K to 
$\sim$\,9000\,K (see the top row of Fig.~\ref{fig:sedopt}) 
and a gradual increase of $R_{\rm WD}$ from $\sim$\,66 to 
$\sim$\,84\ro\ at a constant $L_{\rm WD}$ of 
$\sim$\,1.5$\times 10^{38}$\es. 

About half a day later ($\sim$Aug.~15.5), during the first short 
plateau phase in the light curve ($V\sim 5.6$), the continuum 
profile of 
our spectra reflected an unambiguous increase in the temperature 
to $T_{\rm eff} = 10000\pm 250$\,K, but at nearly unchanged 
radius, $\la 100$\ro, implying an increase of $L_{\rm WD}$ 
by a factor of $\ga 2$. The following heating of the envelope 
to its maximum of $12000\pm 500$\,K on Aug. 15.8--16.0, with 
only a small increase of its radius to 
$\approx 100\,(d/3\,{\rm kpc})$\ro, implied a rapid increase 
of $L_{\rm WD}$ with a maximum of 
$\sim 8.6\times 10^{38}\,(d/3\,{\rm kpc})^2$\es\ on Aug. 16.0 
(see the second row from the top of Fig.~\ref{fig:sedopt}). 

A rapid cooling of the envelope to $\sim$\,7500\,K on Aug. 16.9 
(i.e. at a rate of $\sim$210\,K/hour) was a result of the shell 
expansion by a factor of $\sim$2, to 
$\sim$200\,(d/3\,${\rm kpc})$\ro, which was followed by a general 
decrease of its luminosity to 
$\sim$5$\times 10^{38}\,(d/3\,{\rm kpc})^2$\es. 
At the $V$-brightness maximum (Aug. 16.45, see narrower grey belt 
in Fig.~\ref{fig:lrt}), we indicate a local maximum of 
the luminosity. 

From Aug. 16.9 to the end of the fireball stage of the nova, 
$T_{\rm eff}$ was gradually declining to $\sim$\,6000\,K, 
followed by a gradual increase of its radius up to 
$\approx$\,300$\,(d/3\,{\rm kpc})$\ro, while the corresponding 
luminosity denoted only a small decrease to 
$\sim$\,4.6$\times 10^{38}\,(d/3\,{\rm kpc})^2$\es. 
Our observations caught an interesting behaviour around a shallow 
secondary $V$-band maximum (Aug. 18.4, see broad grey belt in 
Fig.~\ref{fig:lrt}) corresponding to a small local maximum 
of $L_{\rm WD}$. 
%
%
\begin{figure}
\begin{center}
%
\resizebox{8cm}{!}{\includegraphics[angle=-90]{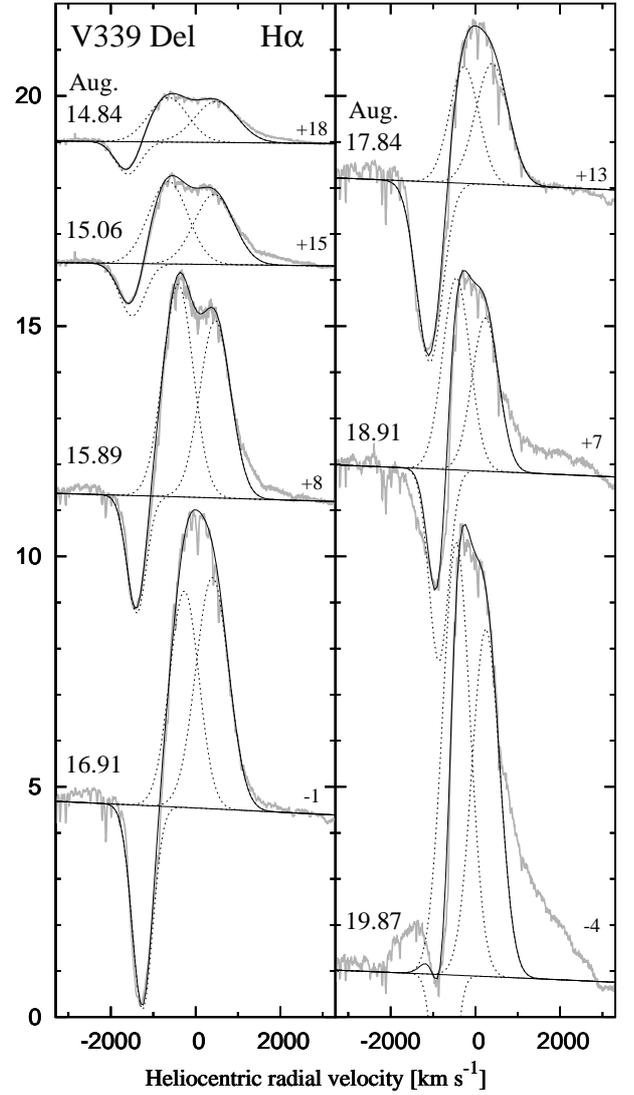}}
\end{center}
\caption[]{
Evolution of the \ha\ line profile during the fireball phase 
of V339~Del (grey lines). Each profile was fitted with three 
Gaussian functions (dotted curves) to isolate its main 
components (Sect. 3.3). 
Numbers on the right side mark the shift of the profile in 
the given units for a better visualization. Dates are marked 
at the left side. Fluxes are in 10$^{-11}$\ecsa. 
          }
\label{fig:ha}
\end{figure}

\subsubsection{Evolution beyond the fireball stage}

Two days after the end of the fireball stage, on Aug.~22 
(day 5.5), the high level of the optical continuum for 
$\lambda > 5000$\,\AA\ (Sect.~3.1) and the bright $JHKLM$ 
magnitudes \citep[][]{cass+13a} required the presence of 
a strong nebular radiation component with 
$\textsl{EM} \sim 2\times 10^{62}\,(d/3\,{\rm kpc})^2$\cmt\ 
supplemented by a contribution from a $\sim 5000$\,K warm 
residual pseudophotosphere scaled to 
$\sim 9\times 10^{37}\,(d/3\,{\rm kpc})^2$\es\ 
(dotted line in Fig.~\ref{fig:sedir}). 
The very high value of \textsl{EM} constrained the minimum of 
$T_{\rm bb} \sim 37000$\,K and 
$L_{\rm WD} \sim 2.2\times 10^{39}\,(d/3\,{\rm kpc})^2$\es\ 
to fit the short-wavelength part of the spectrum 
\citep[for details see Appendix A of ][]{sk15b}. 
The corresponding radius of the WD photosphere was 
$R_{\rm WD} \la 19\,(d/3\,{\rm kpc})$\ro. 

On Aug.~28.5 (day 12), the warm stellar pseudophotosphere 
could no longer be identified.
The spectrum for $\lambda > 6000$\,\AA\ was dominated solely 
by the nebular continuum with 
$\textsl{EM} = 1.8\times 10^{62}\,(d/3\,{\rm kpc})^2$\cmt. 
As in the previous case, the stellar component of radiation 
from the WD photosphere was determined such that its 
Lyman-continuum photons are just capable of giving rise 
to the measured \textsl{EM} and the component fits 
the short-wavelength part of the spectrum. 
In this way, we obtained 
$T_{\rm bb}\ga 43000$\,K, 
$R_{\rm WD} \la 13\,(d/3\,{\rm kpc})$\ro\ and 
$L_{\rm WD}\ga 2.1\times 10^{39}\,(d/3\,{\rm kpc})^2$\es. 

On Sept.~13 (day 28), the nebular 
component of radiation entirely dominated the optical to 
near-IR (330\,nm -- 3.4$\mu$m). A contribution from the WD 
photosphere could not be recognized in the optical spectrum 
(see Figs.~\ref{fig:sedopt} and \ref{fig:sedir}). 
The slope of the continuum and a relatively small Balmer 
jump in emission corresponded to a high $T_{\rm e}$. 
To match the observed continuum we scaled the 
$\varepsilon_{\lambda}(T_{\rm e} = 40000\,{\rm K})$ 
coefficient with $k_{\rm n} = 1.3\times 10^{17}$\,cm$^{-5}$, 
i.e. $\textsl{EM} = 1.4\times 10^{62}(d/3\kpc)^2$\cmt\ 
(Eq.~(\ref{eq:fl2})). 
Because no observations in the UV are available, we can 
only estimate the lower limit of the WD luminosity and 
temperature. There are two principal constraints: 

(i) The WD contribution cannot considerably influence even 
the short-wavelength part of the optical spectrum because 
this would lower the nebular contribution, the amount of 
which in the near-IR is justified by the $HKL$ fluxes. 

(ii) The high \textsl{EM} requires a high rate of 
hydrogen-ionizing photons, 
$L_{\rm ph} = \alpha_{\rm B}(T_{\rm e})\times 
\textsl{EM} \sim 1.4\times 10^{49}$\,s$^{-1}$ 
for the recombination coefficient 
$\alpha_{\rm B}(T_{\rm e}\ge 30000\,K)
\sim 1\times 10^{-13}$\,cm$^{3}$\,s$^{-1}$ 
\citep[e.g.][]{pequignot+91}. 
These conditions thus require the radiation at 
$T_{\rm bb} > 100000$\,K scaled with 
$\theta_{\rm WD} < 1.2\times 10^{-11}$ 
(i.e. $R_{\rm WD} < 1.6$\ro), which correspond to 
$L_{\rm WD} > 8.8\times 10^{38}$\es. 

On Sept.~20 (day 35), the continuum profile remained nearly 
unchanged, but with a flux level by a factor of $\sim$\,1.7 
lower than on Sept.~13 (see Fig.~\ref{fig:sedopt}). 
Taking into account the same conditions as mentioned above, 
the model SED corresponded to $T_{\rm e} \sim 50000\,{\rm K}$, 
$\textsl{EM} = 1.0\times 10^{62}(d/3\kpc)^2$\cmt, which 
constrains $T_{\rm bb} > 150000$\,K, $R_{\rm WD} < 0.6$\ro\ 
and $L_{\rm WD} > 6.5\times 10^{38}$\es. 
However, the near-IR $HKLM$ flux-points from Sept.~21.8 
were located clearly above the nebular continuum 
(see Fig.~\ref{fig:sedir}), indicating the presence of a dust 
component in the spectrum. Our model SED thus confirms the dust 
formation reported by \cite{shenavrin+13} and suggests 
that it occurred already around Sept.~21 
\citep[see also][]{cass+13c}. 
We matched the dust component with a blackbody temperature 
$T_{\rm dust} = 1350\pm 50$\,K and a luminosity 
$L_{\rm dust} = (1.1\pm 0.2)\times 10^{37}$\es. 

Model SEDs during the post-fireball stage are plotted 
in Fig.~\ref{fig:sedir}. 
One of the most interesting results of modelling the SED 
during this period is the persistent super-Eddington 
luminosity, which appeared to be even higher than during 
the fireball phase (see Table~3). 
The same effect was found for RS~Oph by \cite{sk15b}. 

\subsection{Evolution of the \ha\ line profile}

Figure~\ref{fig:ha} shows the evolution of the \ha\ line during 
the fireball stage. Its profile was of P~Cygni-type. To separate 
its main components, we fitted the profile with three Gaussian 
functions, two for its emission component and one to match the 
absorption component (dotted lines in Fig.~\ref{fig:ha}). 
The main characteristics of the temporal evolution of the \ha\ 
profile can be summarized as follows: 

(i)
A strong absorption component was always present. Its radial 
velocity (RV) was gradually decreasing from 
\mbox{$\sim$\,-1600\kms}, as was measured on our first spectrum 
from Aug.~14.84, to $\sim$\,-730\kms, at the maximum of the 
envelope inflation, around Aug.~19.87 
(see panel (a) of Fig.~\ref{fig:hapar}). 
The absorbed flux was gradually increasing until Aug. 18, when 
its amount was by a factor of $\sim$\,8 higher than measured 
on our first spectra (see panel (b) of the figure). 
It is of interest to note that the absorption component
was strong enough to cut parts of the emission component at 
its blue side even after the fireball stage 
($\ga$Aug.~20.0, see Fig.~\ref{fig:hatran}). 

(ii)
Prior to the optical maximum, the emission part of the profile 
was flat-topped, suggesting that it was composed of two strongly 
blended components. After the $V$-maximum, the \ha\ emission 
narrowed, mainly at the blue side, and its core became 
single-peaked. Around Aug. 18, when the absorbed flux reached 
its maximum, a transient decrease in the \ha\ emission was 
observed (panel (c) of Fig.~\ref{fig:hapar}). 

(iii)
At the end of the fireball phase, $\ga$Aug.~18.9, the red emission 
wing expanding to RV$\sim$\,+2750\kms\ increased. Its blue 
counterpart extended to RV$\sim$\,-2550\kms\ since Aug.~19.87. 
These values were almost unchanged during 
the investigated period of the nova evolution. Some examples 
are shown in Fig.~\ref{fig:hatran}. Accordingly, we adopted 
the terminal velocity of the broad wings as their average, 
i.e. $v_{\infty} = 2650\pm 100$\kms. 
%
%
\begin{figure}
\begin{center}
\resizebox{8.0cm}{!}{\includegraphics[angle=-90]{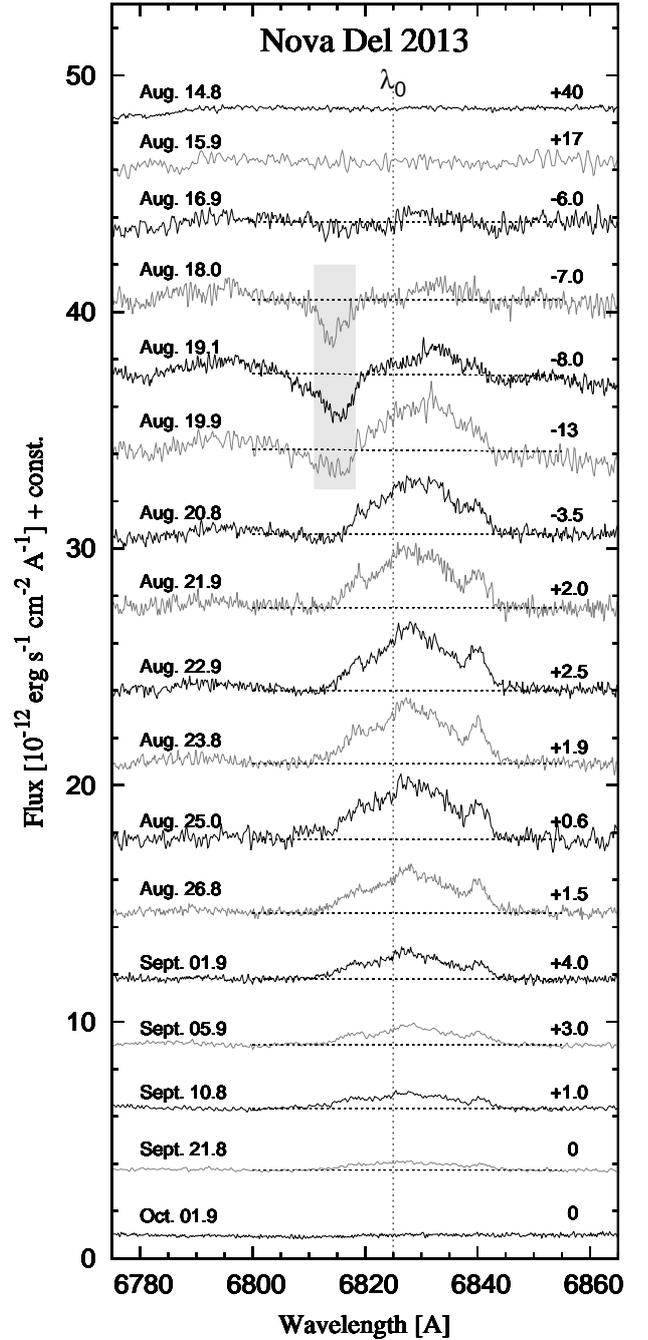}}
%
\end{center}
\caption[]{
Evolution of the Raman-scattered \ion{O}{vi} 1032\,\AA\ line 
becoming evident as an emission feature around $6830$\,\AA\ 
since the onset of the outburst on Aug.~14 to its disappearance. 
The grey belt marks an absorption that developed at the end 
of the fireball stage. 
Numbers on the right side mark the shift of the local 
continuum (dotted lines) with respect to the units of the 
flux scale (arbitrarily added for a better visualization). 
Observing dates are marked at the left side. 
          }
\label{fig:ram1}
\end{figure}
%
%
\begin{figure}
\begin{center}
\resizebox{8.0cm}{!}{\includegraphics[angle=-90]{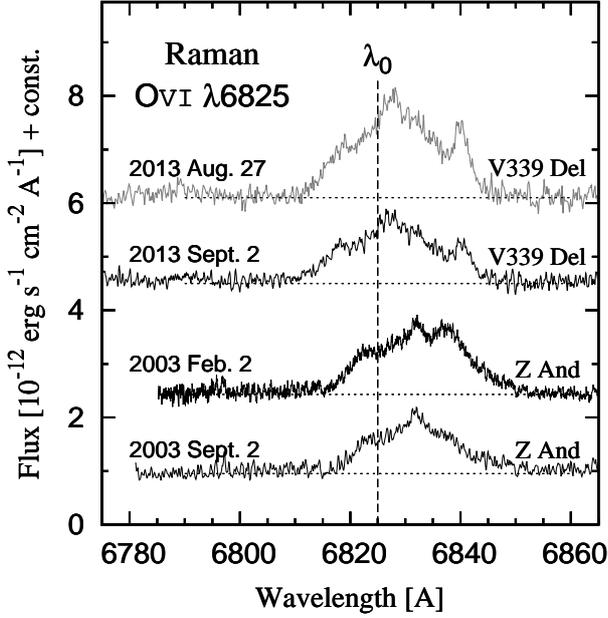}}
%
\end{center}
\caption[]{
Comparison of the Raman 6825\,\AA\ lines from the spectrum 
of V339~Del (the top two profiles) with those observed 
in the spectrum of the classical symbiotic star Z~And at the end 
of its 2000-03 active phase \citep[adapted from][]{skopal+06}. 
The similarity of these profiles supports the presence of 
the Raman-scattering process in the spectrum of the 
{\em classical} nova V339~Del (Sects.~3.4 and 4.2.4). 
          }
\label{fig:ram2}
\end{figure}

\subsection{Raman-scattered \ion{O}{vi} 1032\,\AA\ line -- first 
            time seen in a classical nova}

A transient emergence of the Raman-scattered \ion{O}{vi} 
1032\,\AA\ line in the spectrum of V339~Del was very startling, 
because its induced faint, broad emission feature around 
$\lambda 6830$\,\AA\ has never been observed in the spectrum
of a {\em classical} nova \citep[see][ for the first note]{sk+14}. 

Figure~\ref{fig:ram1} shows the temporal evolution of this feature.
It was observed in the spectrum of V339~Del for more than one month, 
from Aug.~19 to the end of the first plateau phase in the light 
curve around Sept.~23, 2013 (day 37).
During the fireball stage, no Raman emission was detected 
because of the low temperature of the ionizing source, at which 
no original \ion{O}{vi} photons could be created. From Aug.~20 on, 
the flux of the Raman line was decreasing and vanished 
completely by the end of September. 
Its $FWZI$ was $\ga$\,30\,\AA\ and the profile showed a 
redward-shifted component around $6840$\,\AA\ and a faint 
blue-shifted shoulder. A similar structure of the Raman-scattered 
line profile has frequently been recognized in 
the spectra of symbiotic stars (see Fig.~\ref{fig:ram2}). 
Energy conservation of the Raman process implies the broadening 
of the scattered line with a factor of 
$(\lambda_{\rm Ram}/\lambda_{\rm OVI})^2 \sim 44$ 
\citep[][]{nussb+89}. 
Thus, the $FWZI(6830)\approx$\,30\,\AA\ corresponds to a maximum 
broadening of the original 1032\,\AA\ line of $\approx$\,0.7\,\AA, 
i.e. $\approx$\,200\kms, which corresponds to expected movements 
of emitting material in the vicinity of the hot WD. 
Since October~1, just after the first plateau phase of the light 
curve, when a significant X-ray flux was first reported
\citep[][]{page+13}, the Raman line could no longer been 
discerned in our spectra. 

The appearance of the Raman-scattered line during a {\em classical} 
nova outburst is an unexpected important event that places 
strong constraints on the ionization structure of 
the nova ejecta (Sect.~4.2.4). 

\section{Interpretation of observations}

\subsection{Evolution of the parameters $L_{\rm WD}$, 
            $R_{\rm WD}$, and $T_{\rm eff}$}

The comprehensive spectroscopic and photometric observations 
of nova V339~Del, performed immediately after its discovery, 
allowed us to map in detail the evolution of its fundamental 
parameters even during the initial phase of its explosion. 
This showed, however, that the nova was not evolving strictly 
according to the canonical view, which assumes a gradual cooling 
and expanding of the nova envelope at a constant luminosity 
during the fireball stage \citep[e.g.][]{shore+94,schwarz+01}. 

The parameters $L_{\rm WD}$, $R_{\rm WD}$, and $T_{\rm eff}$ 
behaved as commonly expected only after the visual maximum, 
since Aug.~17 (Fig.~\ref{fig:lrt}, Sect.~3.2.1). The evolution 
of the nova prior to the $V$-maximum was unconventional. 
The transient increase in $T_{\rm eff}$ from $\sim 9000$ 
(Aug. 14.97) to $\sim 12000$\,K (Aug. 15.9) at a nearly 
constant $R_{\rm WD}$ was caused by a rapid increase in 
the luminosity from $\sim$\,1.5 to 
$\sim$\,8.6$\times 10^{38}$\es\ (Fig.~\ref{fig:lrt}). 
%
At the maximum of $L_{\rm WD}$ and $T_{\rm eff}$, the effective 
radius $R_{\rm WD}$ started to expand rapidly. During about one 
day, from $\sim$\,Aug.~16 to 17, its value increased by a factor 
of $\ga$\,2, while $T_{\rm eff}$ and $L_{\rm WD}$ decreased and 
followed the standard evolution at approximately constant, but 
super-Eddington luminosity
$L_{\rm WD}\sim5\times 10^{38}\,(d/3\,{\rm kpc})^2$\es, to the 
end of the fireball stage\footnote{
   A small decrease of $L_{\rm WD}$ at the end of the fireball 
   stage was probably only apparent because of the disk-like 
   shape of the pseudophotosphere (see Sect. 4.2.3), which 
   does not allow estimating the total luminosity of the 
   central source \citep[see Sect.~5.3.6 of][]{sk05}.
}. 
An intermediate minimum of $R_{\rm WD}$ around Aug.~17.2 was 
probably caused by a transient decrease of the optical depth 
along the line of sight, resulting in a deeper view into the 
ejecta, down into a hotter optically thick shell, which 
radiates at unchanged $L_{\rm WD}$. On the other hand, a small 
transient increase of $R_{\rm WD}$ around Aug.~18.4 was 
accompanied by an increase in $L_{\rm WD}$ and a small 
decrease of $T_{\rm eff}$ (see Fig.~\ref{fig:lrt}). 
This suggests that the outflowing material was inhomogeneous. 
Another striking result we obtained from our modelling of 
the SED is the very high luminosity of V339~Del, which is well 
above its Eddington limit for $M_{\rm WD} \sim 1.0$\mo\ 
\citep[][]{chochol+14}, during the fireball stage and beyond 
(Table~3). This result also contradicts the theoretical 
expectation that a super-Eddington phase only persists for 
a few hours at the very beginning of the nova eruption 
\citep[e.g.][]{prial+kov95,starrfield+08}. 

Nevertheless, the super-Eddington luminosity was previously 
observationally documented for some novae in the past. 
For example, \cite{friedjung87} showed that the energy 
output of nova FH~Ser was well above the Eddington limit 
for about two months after its optical maximum. 
\cite{schwarz+01} demonstrated that the luminosity of 
the nova LMC~1991 was super-Eddington before its visual 
maximum and reached $L_{\rm WD} \sim 6\times 10^5$\lo. 
Recently, \cite{sk15b} revealed this also for the recurrent 
nova RS~Oph. 

The super-Eddington state was theoretically investigated by 
\cite{shaviv98}, who considered a reduction of the effective 
opacity in the inhomogeneous atmosphere of novae, which 
increases the Eddington luminosity well above its classical 
value, calculated for the Thomson-scattering opacity. 
Lowering the effective opacity can result from the rise 
of a `porous layer' above the convective zone of the burning 
WD \citep[see Fig.~1 of][]{shav+dot10}. Under these conditions 
novae evolve at a super-Eddington steady state, which can 
explain their very long decay times \citep[][]{shav+dot12}. 
The long-term super-Eddington luminosity and the inhomogeneous 
mass ejecta, as we derived from our observations, seem to 
reflect Shaviv's theoretical predictions. 
However, further detailed multiwavelength analyses of other 
objects are required to allow for a more accurate theoretical 
modelling of the nova phenomenon. 

Finally, a very large effective radius of the optically thick 
pseudophotosphere at the end of the fireball stage 
($R_{\rm WD}\sim 300$\ro) suggests a high mass of the 
ejected material (Sect.~4.3). 

\subsection{Biconical ionization structure of the fireball}

\subsubsection{Origin of the \ha\ line}

To understand the origin of the \ha\ line during the fireball 
stage of nova V339~Del, we need to identify (i) the ionizing 
source that is capable of giving rise to the observed \ha\ 
luminosity, and (ii) the material that is subject to ionization, 
the kinematics of which corresponds to the observed RVs. 

\begin{figure}
\begin{center}
\resizebox{8.0cm}{!}{\includegraphics[angle=-90]{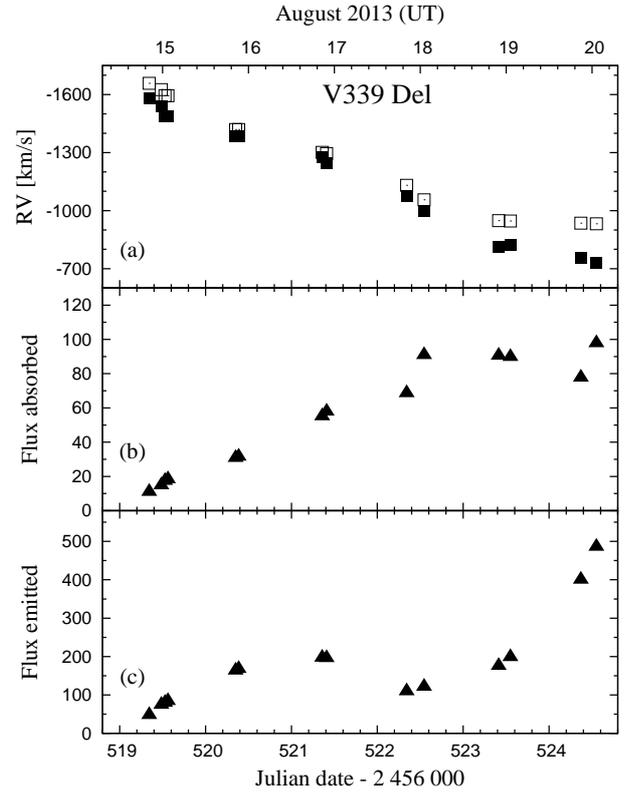}}
\end{center}
\caption[]{ 
Radial velocities and fluxes of the main components of the \ha\ 
profile as a function of time during the fireball stage 
(see Sect.~3.3). Fluxes are in $10^{-11}$\ecs. 
(a) Radial velocities of separated (filled squares) 
    and observed (empty boxes) absorption components. 
(b) Flux absorbed by the \ha\ line. 
(c) Flux emitted by the separated components (from Aug. 18.91 on,
    it also includes the broad emission wing). 
          }
\label{fig:hapar}
\end{figure}

(i) 
Already the first spectra of V339~Del indicated the presence 
of rather strong emission lines of hydrogen. For example, 
the luminosity, $L_{\alpha}$, produced by the \ha\ line 
corrected for the absorption component was between 
5.3$\times 10^{35}$ and 4.1$\times 10^{36}(d/3\kpc)^2$\es, 
as measured on our first (Aug.~14.84) and last (Aug.~19.87) 
spectra during the fireball stage (Fig.~\ref{fig:hapar}). 
This luminosity was generated 
in a volume with the emission measure 
  $\textsl{EM}_{\alpha} = L_{\alpha}/\varepsilon_{\alpha} 
  = 2.9\times 10^{60} - 2.2\times 10^{61}(d/3\kpc)^2$\cmt\ 
for the volume emission coefficient in \ha, 
$\varepsilon_{\alpha}(20000\,{\rm K}) = 1.83\times 
10^{-25}$\,erg\,cm$^{3}$\,s$^{-1}$. 
This requires 
$L_{\rm ph} = \alpha_{\rm B}\times \textsl{EM}_{\alpha} = 
   4.1\times 10^{47} - 3.1\times 10^{48}$ 
ionizations/recombinations per second for the total hydrogen 
recombination coefficient 
  $\alpha_{\rm B}(20000\,{\rm K}) = 
  1.43\times 10^{-13}\,{\rm cm^{3}\,s^{-1}}$ 
\citep[][]{pequignot+91}. 
However, the rate of hydrogen-ionizing photons, produced 
by the 6000--12000 warm pseudophotosphere of the nova, 
$L_{\rm ph}({\rm shell}) = 
2.4\times 10^{41} - 3.1\times 10^{46}$\,photons\,s$^{-1}$, 
is far below the rate of recombinations required to generate 
the $L_{\alpha}$ (i.e. 
$L_{\rm ph}({\rm shell})\ll\alpha_{\rm B}\times EM_{\alpha}$). 
Thus the warm envelope cannot be ascribed to the ionizing source. 
This implies that the ionizing source is not seen directly by 
the observer, because the \textsl{EM}$_{\alpha}$ derived from 
the observed $L_{\alpha}$ is a few orders of magnitude higher 
than what can be produced by the warm pseudophotosphere. 
Such radiative components in the spectrum, which correspond 
to very different and mutually inconsistent temperature regimes, 
are often observed during outbursts of symbiotic stars. 
The corresponding spectrum is called a two-temperature-type 
spectrum \citep[see Sect.~5.3.4 of][]{sk05}. 

\begin{figure}
\begin{center}
\resizebox{8.0cm}{!}{\includegraphics[angle=-90]{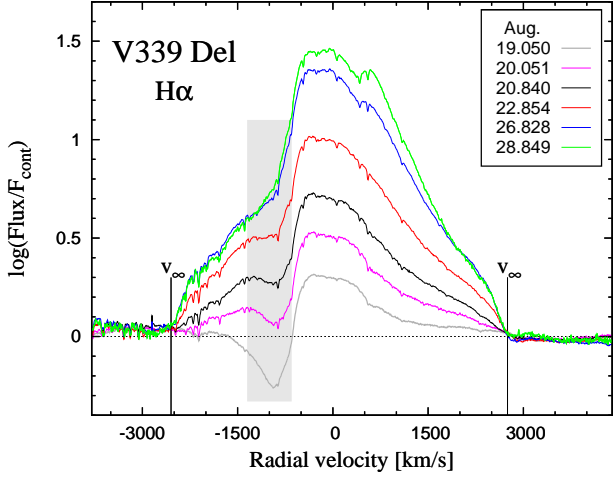}}
\end{center}
\caption[]{ 
Selected \ha\ profiles close to and after the end of the 
fireball stage. The grey belt illustrates the position of 
the absorption component. Terminal velocities of the emission 
wings are denoted by $v_{\infty}$. 
          }
\label{fig:hatran}
\end{figure}

(ii)
An additional constraint for determining the ionizing source 
is provided by the broad \ha\ profile, the red wing of which 
extends to $\ga$\,2000\kms\ during the fireball stage. 
Approaching the end of this phase (after Aug.~18), the broad 
wing has strengthened and extended to a terminal velocity of 
$\sim$\,2700\kms\ (Figs.~\ref{fig:ha} and \ref{fig:hatran}). 
This implies that the mass outflow of the ionized hydrogen must 
have been collimated. Otherwise, the large optically thick 
envelope ($R_{\rm WD} \sim 100-300$\ro) would eclipse 
a significant fraction of the material flowing {\em away} from 
the observer, and would therefore preclude identifying it 
in the form of the extended {\em red} wing of the line. 

\begin{figure}
\begin{center}
\resizebox{8cm}{!}{\includegraphics[angle=-90]{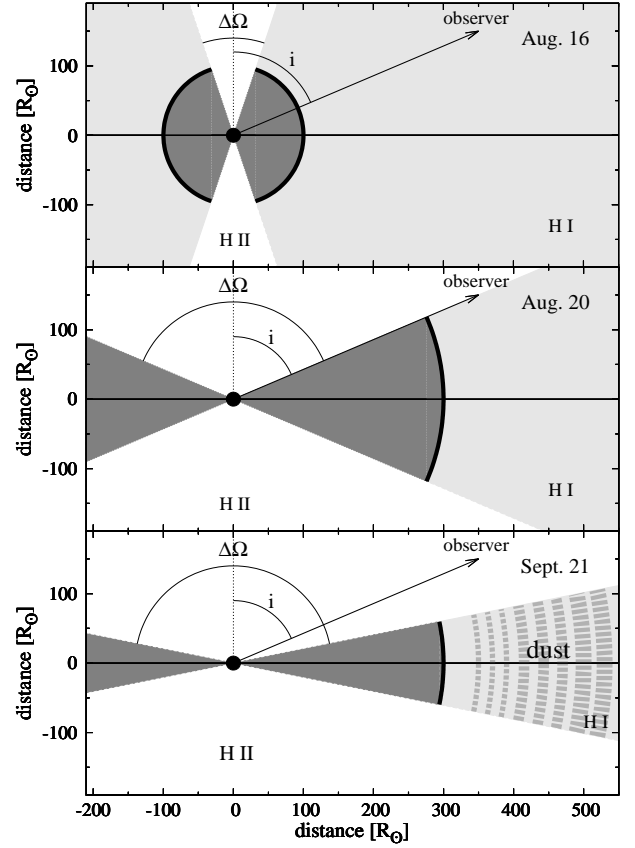}}
%
\end{center}
\caption[]{
Sketch of the ionization structure of nova V339~Del as seen 
on a cut perpendicular to the orbital plane that contains 
the burning WD (the black circle). 
The neutral \ion{H}{i} zone is optically thick to the distance 
of the WD pseudophotosphere (heavy solid line). The ionized 
\ion{H}{ii} zone has the opening angle $\Delta\Omega$, and $i$ 
is the orbital inclination (its value is unknown). 
The top panel shows the structure around the beginning of the 
fireball stage, the middle one at its end, and the bottom 
panel shows the situation when the dust emission is present 
in the ejecta simultaneously with the \ion{H}{i} atoms, hard 
radiation from the burning WD, and the large amount of 
the nebular emission (see Fig.~\ref{fig:sedir}, Sect.~4.2). 
          }
\label{fig:sketch} 
\end{figure}

Observational constraints included in points (i) and (ii) 
above imply that the mass outflow of nova V339~Del was not 
spherically symmetric during the fireball stage, meaning that 
the density was not homogeneously distributed in the envelope. 
Both the flux and the profile of the \ha\ line suggest that 
there is a low-density optically thin part, located in bipolar 
direction relative to the burning WD, probably along its 
rotation axis (i.e. perpendicular to its orbital plane). Its 
particles are ionized by the hot WD surface and driven out 
in the form of a fast wind. The remainder of the surrounding 
envelope is neutral. Its optically thick/thin interface 
represents the warm expanding pseudophotosphere 
(see top panel of Fig.~\ref{fig:sketch}). A similar 
ionization structure of the expanding envelope during the 
fireball phase was also indicated for RS~Oph 
\citep[see Fig.~4 of][]{sk15b}. 
In the case of V339~Del, the opening angle of the \ion{H}{ii} 
zone has to be narrower because the nebular continuum is 
significantly weaker than in the case of RS~Oph. 
%
%
%
%

\subsubsection{Variations of the \ha\ profile}

According to the biconical ionization structure of the envelope 
during the fireball stage, the flat to double-peaked emission 
component of the \ha\ line can be produced by the ionized material 
driven in bipolar directions from the burning WD surface.
Each pole emission can be approximated by one component. 

From Aug.~16 on, a strengthening of the absorption component, 
which reached its maximum around Aug.~18 
(Sect.~3.3, Fig.~\ref{fig:hapar}), was probably caused by the 
inflation of the WD pseudophotosphere by a factor of $>$\,2. 
This was approximately consistent with the enlargement of the 
effective surface of the optically thick envelope (Table~3). 
A deceleration of the envelope expansion caused a gradual 
shift of the absorption component to longer wavelengths 
(see Figs.~\ref{fig:ha} and \ref{fig:hapar}), which narrowed 
the \ha\ emission to a single peak. 
In addition, the significantly expanded pseudophotosphere 
eclipsed a larger part of the \ion{H}{ii} zone, which 
probably caused the transient decrease of its emission as 
observed around Aug.~18 (Fig.~\ref{fig:hapar}). 

The increase of the \ha\ flux and the expansion of the emission 
wings of the line during the last day of the fireball stage 
(Figs.~\ref{fig:ha} and \ref{fig:hatran}) was apparently 
caused by the biconical ``opening'' of the \ion{H}{ii} region 
(see below, Sect.~4.2.3). 
%
%

\subsubsection{Disk-like shape of the envelope and its dilution}

During the last day of the fireball stage (Aug.~19), the emission 
produced by hydrogen lines was clearly increasing 
(see Fig.~\ref{fig:hapar} for \ha). For example, the total flux 
(i.e. corrected for the absorption component) in the \ha\ line 
increased from 1.8 to $3.9\times 10^{-9}$\ecs\ from Aug.~18.91 
to Aug.~19.87. In addition, the nebular continuum also became 
noticeable in our last spectrum taken during the fireball 
stage, on Aug.~19.89 (Fig.~\ref{fig:sedopt}). 
In contrast, $T_{\rm eff}$ and $L_{\rm WD}$ of the warm shell 
were decreasing and were much lower than what would be needed 
to produce the measured nebular emission, as described 
in Sect.~4.2.1. 
This implies that the unseen ionizing source was increasing. 
According to the biconical ionization structure of the expanding 
envelope (see Sect.~4.2.1), the gradual increase of the nebular 
contribution could be caused by a gradual opening of the \ion{H}{ii} 
zone, which would mean a narrowing of the \ion{H}{i} zone. 
The optically thick envelope was therefore developing to 
a disk-like structure. 
Its outer flared rim represented the pseudophotosphere. From 
this point of view, the warm stellar component that dominates 
the spectrum implies that the ionizing source was hidden by 
the outer rim of the disk, that is, it was below the observer's 
horizon represented by the \ion{H}{i}/\ion{H}{ii} boundary. 
As soon as the opening angle of the \ion{H}{ii} zone becomes
$> 2\times i$, the burning WD rises above the boundary and 
thus terminates the fireball stage (see Fig.~\ref{fig:sketch}). 
As a result, we can observe a hotter WD photosphere 
throughout an optically thinner \ion{H}{ii} zone, but a 
contribution from a remnant \ion{H}{i} pseudophotosphere can 
also be expected (see the bottom panel of Fig.~\ref{fig:sketch}). 
This transition was demonstrated by the model SED of Aug.~21.7, 
which indicates a strong nebular component of radiation 
supplemented by a component from the warm residual 
pseudophotosphere and a hotter WD photosphere 
(Sect.~3.2.2, Fig.~\ref{fig:sedir}). 
In the $V$-light curve this was reflected by the first rapid 
fading from a flat maximum on Aug.~19 
\citep[see Fig.~2 of][]{munari+13a}.  

A significant decrease of $R_{\rm WD}$ from $\sim 300$\ro\ to 
$\le 19$\ro\ in two days is consistent with a rapid decrease 
of the optical depth along the line of sight caused mainly by 
the geometrical effect of the rise of the hot WD above the 
\ion{H}{i}/\ion{H}{ii} horizon (see Fig.~\ref{fig:sketch}). 
A similar effect of a sudden shift of the SED maximum to 
the ultraviolet was also indicated in the spectrum of nova 
RS~Oph, which occurred during a much shorter time than 
the corresponding dynamical time-scale 
\citep[see Sect.~4.2 of][]{sk15b}. 

A gradual opening of the \ion{H}{ii} zone can be a result of 
a decreasing mass-loss rate from the WD (see Sect.~4.4). 
Recently, \cite{cask12} showed that the biconical ionization 
structure can be formed as a consequence of the enhanced 
mass-loss rate from the rotating WD during outbursts of symbiotic 
binaries. In their model, the opening of the \ion{H}{ii} zone 
is related to the mass loss rate from the WD as 
$\dot M_{\rm WD}^{-2}$ (see their Figs. 1 and 6). 

Finally, there are two consequences of this type of the nova 
envelope dilution. 

(i) 
The duration of the fireball stage can be a function of $i$. 
In general, a higher inclination implies a longer-lasting 
fireball stage, and vice versa. The transition of the maximum 
of the SED from the optical to the UV can occur during a short 
time of a few days. 

(ii) 
The neutral material within the \ion{H}{i} zone, which has the 
form of a disk encompassing the hot WD, can persist within the 
ejecta for a much longer time than the duration of the fireball 
phase until it becomes totally ionized. It is evident as 
an extra component in the model SEDs during a few days 
after the fireball stage (Fig.~\ref{fig:sedir}, Sect.~3.2.2). 
Further evident proof of the \ion{H}{i} zone beyond the fireball 
stage is the Raman-scattered \ion{O}{vi} 1032\,\AA\ line in 
the optical spectrum for more than one month (Sect.~3.4). 

\subsubsection{Probing the neutral disk by the Raman-scattering}

In the process of Raman-scattering, a photon excites an atom 
from its ground state to an {\rm intermediate} state, which is 
immediately stabilized by a transition to a different true 
bound state. This process produces the Raman-scattered photon 
at a (very) different frequency and the corresponding rest line 
photon \citep[see Fig.~1 of][]{nussb+89}. 

The Raman-scattering of the \ion{O}{vi} 1032\,\AA\ line photons 
by neutral hydrogen atoms produces a broad emission band 
at 6825\,\AA. This Raman conversion was clearly indicated in the 
spectra of many symbiotic stars. Its efficiency depends on the 
scattering geometry; more accurately, on the fraction of the sky 
as seen from the \ion{O}{vi} zone, which is covered by the neutral 
gas where Raman scattering can take place. 
A small cross-section of the 
 $\lambda1032 \rightarrow \lambda6825$ Raman conversion, 
 $\sigma_{\rm Ram} \sim 4.4\times 10^{-24}$\cmdd\ 
\citep[][]{schmid89} requires a large amount of neutral material 
in the circumstellar environment where it originates. 
The column density of the \ion{H}{i} atoms along the line of sight 
of the incident \ion{O}{vi} photons must exceed a value of 
$N_{\rm H} \ga 10^{23}$\cmd to produce an observable effect 
(i.e., the optical depth $N_{\rm H}\sigma_{\rm Ram}\ga 1)$. 

From the properties of the Raman-scattering process and its 
induced feature observed around 6830\,\AA\ (see Fig.~\ref{fig:ram1}, 
Sect.~3.4) we can infer some characteristics of both the neutral 
\ion{H}{i} and the ionized \ion{O}{vi} zone. 

(i) 
The Raman-scattered \ion{O}{vi} 1032\,\AA\ line in the spectrum 
of V339~Del confirms that there is a massive \ion{H}{i} zone in 
a classical nova as well. This seems surprising, since there is 
no other source of neutral hydrogen, as there is in symbiotic 
binaries. However, according to the ionization structure of 
the envelope and its evolution during and after the fireball 
stage (Sects.~4.2.1 and 4.2.3), the Raman-scattering takes 
place in the neutral disk-like formation that encompasses 
the burning WD (see also Fig.~\ref{fig:sketch}). 

(ii) 
The profile of the Raman line did not change markedly 
throughout its visibility. This implies that the {\em geometry} 
of the scattering \ion{H}{i} region (i.e. the disk around 
the hot ionizing source) also persisted more or less unchanged. 

(iii) 
A gradual decrease of the Raman flux since the end of August 
reflected a gradual decrease of the Raman-scattering efficiency 
due to a narrowing of the disk-like \ion{H}{i} zone; as a 
consequence, the solid angle of sky area filled with neutral 
gas as seen from the \ion{O}{vi} photon-emitting zone becomes 
smaller. This is probably a result of a decreasing mass-loss 
rate (see Sect.~4.4), which leads to a gradual ionization and
hence flattening of the neutral disk \citep[][]{cask12}, 
which in turn leads to an enlargement of the \ion{H}{ii} 
zone (see Sect.~4.3). 

(iv) 
The structure of the Raman profile itself results from the 
relative motions between the main emitting and scattering 
regions. A redward shift of the Raman-scattered emission 
is consistent with its origin predominantly in the neutral 
disk that expands from the hot WD. A similar interpretation was 
suggested for the symbiotic prototype Z~And during its 2006 
post-outburst stage \citep[see Fig.~5 of][]{skopal+09}. 

(v) 
Finally, the Raman-scattered \ion{O}{vi} 1032\,\AA\ line in 
the spectrum requires the existence of an ionizing source 
that is capable of producing the O$^{+5}$ ions. According to 
\cite{mn94}, the ionization potential $\chi$(O$^{+5}) \sim 114$\,eV 
requires $T_{\rm WD} \sim 114\,000$\,K. At this temperature, other 
permitted lines with lower ionization potential (e.g. \ion{He}{ii} 
lines) are also expected in the spectrum. However, no 
\ion{He}{ii} lines were identified at the same time as the 
Raman 6830\,\AA\ emission. This is a puzzle. We can only speculate 
that the emission regions with highly ionized elements are very 
small \citep[an example is given in Appendix B of][]{skopal+06}, 
so that the surrounding dense layers can absorb photons of 
corresponding transitions. However, especially the \ion{O}{vi} 
1032\,\AA\ photons can be effectively absorbed by Rayleigh and 
Raman scattering. The former is a multiple-scattering resonance 
process (it generates a photon of the same frequency), which can 
be terminated by the latter one -- the last transition in the 
Rayleigh-scattering chain, which produces the Raman 6825\,\AA\ 
photon. As there is no other strong transition of any other 
element at/around this wavelength \citep[e.g.][]{allen80}, 
the Raman photon can easily escape from the region where 
it was created. 

\subsection{Mass of the emitting material}

According to our modelling results for the SED, the nebular 
emission in the continuum arose at the end of the fireball 
stage, when the broad wings of the \ha\ line also expanded to 
$v_{\infty} = \pm 2650$\kms\ (Sect.~3.3, Fig.~\ref{fig:hatran}). 
Assuming that the hydrogen recombination lines are produced 
within the same emission region as the nebular continuum, its 
volume can be estimated as 
%
%
\begin{equation}
 V_{\rm neb} = \epsilon \frac{4}{3} \pi R_{\rm neb}^3 ,
\label{eq:vem}
\end{equation}
where the filling factor $\epsilon < 1$ reduces the spherical 
volume by taking into account the biconical structure of the 
\ion{H}{ii} zone due to the neutral disk that encompasses 
the WD (see Fig.~\ref{fig:sketch}). 
In addition, the disk blocks a fraction of the ionized 
region from its opposite side in the direction of the 
observer. The largest radius of the emitting volume is 
$R_{\rm neb} = v_{\infty}\times t$, where $t$ is the time 
elapsed since the end of the fireball stage (Aug.~19.9, 2013). 
Then, according to the definition of the emission measure, 
the average particle density of the nebular-emitting region 
can be approximated as 
   $\bar n = (\textsl{EM}/V_{\rm neb})^{1/2}$. 
With the aid of these simple relations, we can write 
the emitting mass of the nebula, 
$M_{\rm neb} = \mu m_{\rm H}\,\bar{n}\,V_{\rm neb}$, 
in the form 
\begin{equation}
M_{\rm neb} = \xi \times \epsilon^{1/2} 
              \left(\frac{EM}{[{\rm cm^{-3}}]}\right)^{1/2} 
              \left(\frac{v_{\infty}}{[{\rm km\,s^{-1}}]} 
                    \frac{t}{[{\rm d}]}\right)^{3/2}~
                    M_{\sun}, 
\label{eq:mem}
\end{equation}
where the factor $\xi$ = $1.9\times 10^{-42}$ for the mean 
molecular weight $\mu = 1.4$ and the mass of the hydrogen 
atom $m_{\rm H} = 1.67\times 10^{-24}$\,g. 

On Aug.~22 ($t\sim 2$\,days), the model 
$\textsl{EM} = 2\times 10^{62}(d/3\kpc)^2$\cmt\ corresponded 
to a maximum ($\epsilon =1$) mass of the emitting material, 
$M_{\rm neb} \sim 1\times 10^{-5}$\mo. The subsequent quantities 
of $M_{\rm neb}$ derived for Aug.~28 ($t\sim 9$\,days), 
Sept.~13 ($t\sim$\,25\,days), and Sept.~20 ($t\sim$\,31.5\,days) 
were increasing to $9.3\times 10^{-5}$, $3.8\times 10^{-4}$, 
and $4.6\times 10^{-4}$\mo, respectively (for \textsl{EM} 
in Table~3). 
Such a large increase of $M_{\rm neb}$ could not only be caused 
by a high mass-loss rate from the burning WD (see below), 
but mainly by a gradual dilution (i.e. ionization) of the neutral 
disk-like material, as indicated by the evolution of the Raman 
6825\,\AA\ \ion{O}{vi} line (Fig.~\ref{fig:ram1}). 

Considering only the biconical shape of the emitting region, 
the filling factor $\epsilon = 2\Delta\Omega/4\pi$, where 
$\Delta\Omega$ is the opening angle of the \ion{H}{ii} region 
in sr. For example, $\Delta\Omega = \pi$ gives $\epsilon = 0.5$, 
which lowers $M_{\rm neb}$ by a factor of 0.7. 
The very high \textsl{EM} suggests a large $\Delta\Omega$ 
with $\epsilon$ close to 1, and thus no significant 
reduction of $M_{\rm neb}$ due to the filling factor in 
Eq.~(\ref{eq:mem}) can be expected. 
%

\subsection{Mass-loss rate of the ionized material}

After the fireball stage, the nebular component of radiation 
dominated the optical to near-IR spectrum of V339~Del 
(Fig.~\ref{fig:sedir}). It represents a fraction of the WD's 
radiation re-processed by ionization processes followed by 
recombinations and free--free transitions. 
Thus, knowing the physical process and its result, the 
\textsl{EM} from the model SED allows us to determine the 
mass-loss rate of the ionized material, $\dot M_{\rm WD}$. 

In our simplified approach the material flows out at a constant 
velocity $v_{\infty}$, and its density distribution $n(r)$ at 
a radial distance $r$ from its source satisfies the mass 
continuity equation as 
\begin{equation}
 \dot M_{\rm WD} = 2 \Delta\Omega\,r^2\,\mu m_{\rm H}\,
                    n(r) v_{\infty}, 
\label{eq:mdot}
\end{equation}
where $\Delta\Omega < 2\pi$ is the opening angle of the 
\ion{H}{ii} region. It is further assumed that all particles 
become ionized on their path from the place of origin, 
$R_{\rm WD}$, to the end of the nebula, $R_{\rm neb}$. 
Then, in the case of a pure hydrogen gas, we can approximate 
the equilibrium condition between the flux of ionizing 
photons, $L_{\rm ph}$, and the rate of recombinations 
in the nebula as 
\begin{equation}
  L_{\rm ph}({\rm H})\, =\, 
               \alpha_{\rm B}({\rm H},T_{\rm e})\!
               \int_{\rm HII}\!\!n_{p}(r)n_{\rm e}(r)
               \,{\rm d}V,
\label{eq:lph}
\end{equation}
where $\alpha_{\rm B}({\rm H},T_{\rm e})$ (cm$^{3}$\,s$^{-1}$) 
is the recombination coefficient to all but the ground state of 
hydrogen \citep[see also Eq.~(4) of][]{nussvog87}. This equation 
is valid for a hydrogen plasma heated by photoionizations 
and characterized by a constant $T_{\rm e}$. 
According to the geometry of the \ion{H}{ii} region, 
${\rm d}V = 2 \Delta\Omega r^2 {\rm d}r$. For a complete 
ionization ($n_{\rm e}(r) = n_{\rm p}(r) = n(r))$, the equilibrium 
equation (\ref{eq:lph}) can be integrated from $R_{\rm WD}$ to 
$R_{\rm neb}$, and using $n(r)$ given by Eq.~(\ref{eq:mdot}), 
$\dot M_{\rm WD}$ can be expressed as 
%
\begin{equation}
 \dot M_{\rm WD} = 
         \left[2\Delta\Omega\,(\mu m_{\rm H} v_{\infty})^2 
         \frac{L_{\rm ph}({\rm H})}
              {\alpha_{\rm B}({\rm H},T_{\rm e})}
         \left(\frac{1}{R_{\rm WD}} - \frac{1}{R_{\rm neb}}
         \right)^{-1}\right]^{1/2}  {\rm g s^{-1}},
\label{eq:mdot2}
\end{equation}
where the ratio 
$L_{\rm ph}({\rm H})/\alpha_{\rm B}({\rm H},T_{\rm e})$ = 
\textsl{EM}. 

On Aug.~22, the parameters of the model SED, 
$R_{\rm WD} = 19$\ro, $R_{\rm neb} = 4.6\times 10^{13}$\,cm
and $\textsl{EM} = 2\times 10^{62}(d/3\kpc)^2$\cmt\ 
yield $\dot M_{\rm WD} = 5.7\times 10^{-4}$\myr. 
On Aug.~28, Sept.~13, and 20, $\dot M_{\rm WD}$ was 
decreasing to 4.5, 1.4, and $0.71\times 10^{-4}$\myr, 
respectively (parameters in Table~3). In all cases we adopted 
$2\Delta\Omega\ = 4\pi$ and $v_{\infty} = 2650$\kms\ 
(Sect.~3.3), which correspond to maximum values of 
$\dot M_{\rm WD}$. 

\section{Summary}

We modelled the optical/near-IR SED of the classical nova 
V339~Del from its discovery on Aug.~14, 2013 until the end of 
the first plateau phase in the light curve (around day 40). 
We determined the fundamental parameters $L_{\rm WD}$, 
$R_{\rm WD}$, and $T_{\rm eff}$ of the burning WD 
pseudopotosphere (Fig.~\ref{fig:lrt}). Monitoring the evolution 
of the \ha\ line profile and the transient emergence of 
the Raman-scattered 1032\,\AA\ line in the optical 
spectrum allowed us to determine the ionization structure of 
the nova during this early period of its evolution 
(Fig.~\ref{fig:sketch}). 
The main results of our analysis may be summarized as follows. 
\begin{enumerate}
\item
During the fireball stage (Aug. 14.84--19.89, 2013; see Sect.~3.1), 
$T_{\rm eff}$ was in the range of 6000--12000\,K, $R_{\rm WD}$ 
was expanding in a non-uniform way from $\sim$\,66 to 
$\sim\,300\,(d/3\,{\rm kpc})$\ro\ and $L_{\rm WD}$ was 
super-Eddington, but not constant. Its maximum occurred around 
1.5 day after the nova discovery, at the maximum of $T_{\rm eff}$ 
(Aug. 16.0), when $L_{\rm WD}$ rapidly increased by a factor of 
$\sim$\,6 with respect to its initial value of 
$\sim$\,$1.5\times 10^{38}\,(d/3\,{\rm kpc})^2$\es\ 
(Sect.~3.2.1, Fig.~\ref{fig:lrt}). 
Thus, V339~Del did not obey the theoretical prediction that 
a nova evolves at a constant bolometric luminosity in the early 
stages of outburst (Sect. 4.1). 
\item
After the fireball stage ($\ga$Aug.~20), the optical/near-IR SED 
changed significantly (Figs.~\ref{fig:sedopt} and \ref{fig:sedir}). 
The continuum was dominated by the nebular radiation component 
with a large 
$\textsl{EM} = 1.0-2.0\times 10^{62}\,(d/3\,{\rm kpc})^2$\cmt. 
As a result, only limiting values of the $L,R,T$ parameters 
could be determined from the measured \textsl{EM} and the profile 
of the SED (Sect.~3.2.2). The lower limit of $L_{\rm WD}$ 
was still super-Eddington, being around 
$1\times 10^{39}\,(d/3\,{\rm kpc})^2$\es, which is higher than 
the value we derived during the fireball stage. 
\item
The profile of the \ha\ line and its relative high flux even 
during the fireball phase imply that there is a low-density 
optically thin part of the envelope, with a bipolar shape 
relative to the burning WD. This part of the envelope is 
ionized, while the remainder is neutral. The optically 
thick/thin interface represents the warm expanding 
pseudophotosphere. 
A gradual increase of the nebular contribution during the last 
day of the fireball stage (Aug. 18.9 -- 19.9) could be caused by 
a gradual opening and enlargement of the \ion{H}{ii} zone, 
causing a narrowing of the \ion{H}{i} zone. The optically thick 
pseudophotosphere was therefore transformed into a disk-like shape 
(Sect.~4.2, Fig.~\ref{fig:sketch}). 
\item
According to the biconical ionization structure of the nova ejecta, 
the rapid decrease of $R_{\rm WD}$ from $\sim$\,300 to $\sim$\,19\ro\ 
and the dramatic change in the SED during a few days after the 
fireball stage was caused by the rising of the hot WD above the 
\ion{H}{i}/\ion{H}{ii} horizon with respect to the line of 
sight (Fig.~\ref{fig:sedir}, Sect. 4.2.3). 
\item
A disk-like \ion{H}{i} region was indicated within the ejecta 
for a period of $\ga$\,1 month after the fireball stage, until 
its total ionization, around day 40. 
The most conclusive proof of the presence of the \ion{H}{i} zone 
around the burning WD is the presence of the Raman-scattered 
\ion{O}{vi} 6830\,\AA\ emission, which requires a column density 
of the \ion{H}{i} atoms of $N_{\rm H} \gtrapprox 10^{23}$\cmd\ 
(Sect.~4.2.4, Fig.~\ref{fig:ram1}). 
\item
On Sept.~20, our model SED indicated a dust emission in the near-IR. 
The simultaneous presence of the hard radiation from the burning WD 
constrains the dust to be located in a ring beyond the \ion{H}{i} 
zone, where it is shielded from the high-temperature radiation 
(bottom panel of Fig.~\ref{fig:sketch}). 
\item
The emitting mass of the \ion{H}{ii} zone was gradually growing 
from $\sim 1\times 10^{-5}$ (Aug.~22) to 
$\sim 4.6\times 10^{-4}$\mo\ (Sept.~20) mainly due to a gradual 
ionization of the \ion{H}{i} zone (Sect.~4.3). 
\item
The mass-loss rate from the nova was decreasing from 
$\sim$\,5.7$\times 10^{-4}$ on Aug.~22 to 
$\sim$\,7.1$\times 10^{-5}$\myr\ on Sept.~20 (Sect.~4.4). 
\end{enumerate}
As in the case of the extraordinary classical nova LMC~1991
\citep[][]{schwarz+01} and/or the symbiotic recurrent nova 
RS~Oph \citep[][]{sk15b}, the special physical conditions 
derived from observations of the classical nova V339~Del 
represent new challenges for the theoretical modelling of 
the nova phenomenon. 

\begin{acknowledgements}
We thank the anonymous referee for constructive comments. 
The spectra presented in this paper were in part obtained 
within the {\it Astronomical Ring for Access to Spectroscopy 
(ARAS)}, an initiative promoting cooperation between 
professional and amateur astronomers in the field of spectroscopy. 
The authors thank ARAS observers for their contributions made 
within the ARAS programme coordinated by Francois Teyssier. 
We also acknowledge the variable-star observations 
from the AAVSO International Database contributed by 
observers worldwide and used in this research. 
This research has been in part supported by the project 
No. SLA/103115 of the Alexander von Humboldt foundation 
and by a grant of the Slovak Academy of Sciences VEGA 
No.~2/0002/13. 
\end{acknowledgements}

\end{document}